\documentclass[%
 reprint,
 amsmath,amssymb,
 aps,
 prl,
]{revtex4-2}

\usepackage{natbib}
\usepackage{graphicx}% Include figure files
\usepackage{dcolumn}% Align table columns on decimal point
\usepackage{bm}% bold math
\usepackage{physics}
\usepackage{tikz}
\usepackage{amsmath}
\usepackage{amsthm}
\usepackage{amssymb}
\usepackage{algorithm}
\usepackage{qcircuit}
\usepackage{physics}
\usepackage{tikz}
\usepackage{graphicx}
\usepackage{hyperref}
\newtheorem{definition}{Definition}
\newtheorem{lemma}{Lemma}
\newtheorem{theorem}{Theorem}
\newtheorem{corollary}{Corollary}

\begin{document}

\preprint{APS/123-QED}

\title{Constant-Depth Quantum Circuits for Arbitrary Quantum State Preparation via Measurement and Feedback}

\author{Wei Zi}
\thanks{ziwei20z@ict.ac.cn}
\affiliation{State Key Lab of Processors, Institute of Computing Technology, Chinese Academy of Sciences, Beijing 100190, China}
\affiliation{School of Computer Science and Technology, University of Chinese Academy of Sciences, Beijing 100049, China}
\author{Junhong Nie}
\thanks{niejunhong19z@ict.ac.cn}
\affiliation{State Key Lab of Processors, Institute of Computing Technology, Chinese Academy of Sciences, Beijing 100190, China}
\affiliation{School of Computer Science and Technology, University of Chinese Academy of Sciences, Beijing 100049, China}
\author{Xiaoming Sun}
\thanks{sunxiaoming@ict.ac.cn}
\affiliation{State Key Lab of Processors, Institute of Computing Technology, Chinese Academy of Sciences, Beijing 100190, China}
\affiliation{School of Computer Science and Technology, University of Chinese Academy of Sciences, Beijing 100049, China}

\begin{abstract}
The optimization of quantum circuit depth is crucial for practical quantum computing, as limited coherence times and error-prone operations constrain executable algorithms. Measurement and feedback operations are fundamental in quantum computing (e.g., quantum error correction); we develop a framework using them to achieve constant-depth implementations of essential quantum tasks. This includes preparing arbitrary quantum states with constant-depth circuits through measurement and feedback, breaking the linear-depth lower bound that is required without these operations. Our result paves the way for general quantum circuit compression using measurement and feedback.

% Measurement and feedback play a pivotal role in the realms of quantum error correction and numerous specialized quantum algorithms. In this manuscript, we establish a systematic approach for preparing arbitrary quantum states with constant-depth circuits through measurement and feedback. 
% Our result can be generalized to controlled quantum state preparation and an entangled version of unitary synthesis, both achievable through constant-depth quantum circuits with the aid of measurement and feedback.
% Moreover, we conjecture these implementations achieve simultaneous optimality in both circuit depth and size for the three problems.
% Our result paves the way for executing more extensive quantum computations within a minimal circuit depth.

% Measurement and feedback mechanisms play a pivotal role in the realms of quantum error correction and numerous specialized quantum algorithms. In this paper, we propose a methodology that uses measurement and feedback to perform general quantum computing tasks using only constant-depth circuits. Examples of such tasks include quantum state preparation (QSP) and an entangled variant of unitary synthesis. The circuit size nearly reaches the lower bound for both QSP and unitary synthesis. This paper paves the way for executing more extensive quantum computations within a minimal circuit depth.

\end{abstract}

\maketitle

\section{Introduction}

Quantum computation has been theoretically proven to outperform classical computation in solving specific problems \cite{shor1994algorithms,grover1996fast}. In recent years, significant advancements have been achieved in the development of quantum devices \cite{google2024quantum}. However, the available quantum computational resources remain severely limited. A major challenge is decoherence \cite{mazzola2010sudden}, which necessitates the completion of quantum computations as quickly as possible to ensure result reliability. Consequently, optimizing the depth of quantum circuits has become a critical and highly significant research focus \cite{baumer2024quantum,allcock2023constant,jiang2020optimal,zi2025shallow,takahashi2016collapse,baumer2024measurement,baumer2024measurement}.

Measurement and feedback are fundamental and indispensable operations in quantum circuit execution \cite{carrera2024combining}. They play a critical and independent role in quantum error correction \cite{google2024quantum} and can significantly accelerate various quantum computation tasks. Examples include quantum fan-out \cite{baumer2024measurement}, long-range CNOT gates \cite{baumer2024efficient}, $n$-Toffoli gates \cite{takahashi2016collapse}, Quantum Fourier transforms \cite{baumer2024quantum}, Uniformly Controlled Gates \cite{allcock2023constant}, and numerous specific types of quantum state preparation \cite{buhrman2024state,piroli2024approximating,lu2022measurement,tantivasadakarn2023hierarchy,lu2023mixed,smith2024constant,tantivasadakarn2023shortest,smith2023deterministic,zhu2023nishimori,malz2024preparation,stephen2024preparing,zhang2024characterizing,sahay2024finite,piroli2021quantum}. In recent years, significant progress has been made in the physical implementation of measurement and feedback \cite{iqbal2024topological,wan2019quantum,corcoles2021exploiting,foss2023experimental}. In particular, recent experimental breakthroughs in quantum error correction \cite{google2024quantum} have further demonstrated the reliability of measurement feedback as a fundamental operation in quantum circuits. While measurement and feedback have been successfully applied to accelerate specific computation tasks, their potential for addressing more general computation problems remains an open and intriguing area of research.

Quantum state preparation (QSP) has garnered significant attention in recent years \cite{PhysRevLett.129.230504,sun2023asymptotically,yuan2023optimal}, owing to its critical applications in quantum machine learning \cite{harrow2009quantum,lloyd2014quantum,kerenidis2016quantum} and Hamiltonian simulation \cite{childs2018toward,low2019hamiltonian}. In this work, we demonstrate that an arbitrary quantum state can be prepared using a constant-depth circuit consisting of 31 layers of intermediate measurements. Additionally, we show that the circuit size and ancilla count can be optimized at the cost of increasing the circuit depth, while maintaining the constant-depth property. A summary of these results is presented in Table~\ref{tab:result}.

%When aiming to optimize both the circuit depth and size concurrently, we show that for an arbitrary quantum state, it is feasible to construct a circuit capable of preparing it. The size of this circuit nearly reaches the lower bound for QSP in the absence of measurement and feedback \cite{plesch2011quantum}. Additionally, this circuit has a constant depth with a maximum of 81 layers of intermediate measurements. Our results suggest that they are likely to be asymptotically optimal, considering not only the circuit depth but also the circuit size.

\begin{table*}
    \centering
    \caption{A summary of our results is presented below, where $n$ denotes the number of qubits. We show that all the following problems can be solved using a constant-depth quantum circuit, augmented by intermediate measurements and feedback.}
    \renewcommand{\arraystretch}{1.5}
    \begin{tabular}{c|c|c|c|c}
    \hline \hline
       Problem  & circuit depth & layers of measurements & circuit size & ancilla count \\  \hline
       Quantum state preparation  & $O(1)$ & $31$ & $O(n4^n)$ & $O(n4^n)$ \\ 
       Quantum state preparation  & $O(1)$ & $81$ & $O(n2^n)$ & $O(n2^n)$ \\ 
       Quantum state preparation  & $O(1)$ & $O(1)$ & $\approx O(2^n)$ & $\approx O(2^n)$ \\ 
       Controlled quantum state preparation  & $O(1)$ & $50$ & $O(n4^n)$ & $O(n4^n)$ \\ 
       Controlled quantum state preparation  & $O(1)$ & $O(1)$ & $\approx O(4^n)$ & $\approx O(4^n)$ \\ 
       Entangled unitary synthesis & $O(1)$ & $50$ & $O(n4^n)$ & $O(n4^n)$ \\
       Entangled unitary synthesis & $O(1)$ & $O(1)$ & $\approx O(4^n)$ & $\approx O(4^n)$ \\ \hline \hline
    \end{tabular}
    \label{tab:result}
\end{table*}

We further investigate an entangled variant of unitary synthesis. Traditional unitary synthesis aims to construct a circuit \( C \) that transforms a quantum state \( \sum_j \alpha_j \ket{j} \) (where \( \alpha_j \in \mathbb{C} \)) into \( \sum_j \alpha_j U \ket{j} \) \cite{vartiainen2004efficient}. In contrast, our work focuses on constructing a circuit that transforms a quantum state \( \sum_j \alpha_j \ket{j} \otimes \ket{0}^{\otimes n} \) into \( \sum_j \alpha_j \ket{j} \otimes U \ket{j} \). We define this process as "entangled unitary synthesis."
In many scenarios, entangled unitary synthesis can be viewed as analogous to traditional unitary synthesis. For instance, this is the case when the initial state is a computational basis state. We demonstrate that entangled unitary synthesis can be achieved using a constant-depth circuit with 50 layers of intermediate measurements. This result is derived from our ability to implement controlled quantum state preparation with a constant-depth circuit of 50 layers of intermediate measurements. Specifically, our objective is to prepare an \( n \)-qubit quantum state controlled by \( n \) control qubits. The results are summarized in Table~\ref{tab:result}.

% Simultaneously, the size of this circuit nearly matches the lower bound of unitary synthesis \cite{shende2004minimal}. Our results indicate the potential to accelerate more general quantum computation missions through the utilization of measurement and feedback.

It is worth noting that, in the absence of measurements and feedback, the size of a quantum circuit for state preparation has a lower bound of $\Omega(2^n)$ \cite{plesch2011quantum}, while circuits for unitary synthesis and controlled state preparation have a lower bound of $\Omega(4^n)$ \cite{shende2004minimal,yuan2023optimal}. Based on these lower bounds, we conjecture that our results simultaneously achieve both constant circuit depth and asymptotically optimal circuit size.

We present several additional results enabled by measurements and feedback. First, we prove that an arbitrary reversible function oracle can be implemented using a constant-depth circuit with $18$ layers of intermediate measurements. Reversible functions can represent all feasible classical computations within the quantum circuit model \cite{wu2024asymptotically}. Second, we demonstrate that both the GHZ state and the quantum fan-out gate can be implemented using constant-depth circuits. These circuits require only one layer of intermediate measurements and $n/c$ ancilla, where $c$ is an arbitrary positive integer. Furthermore, we extend this result to qudit systems.

The paper is organized as follows. In Section~\ref{sec:pre}, we introduce the necessary preliminaries. In Section~\ref{sec:qsp}, we present the construction of a constant-depth quantum circuit for arbitrary quantum state preparation. In Section~\ref{sec:cqsp}, we generalize our approach to controlled quantum state preparation and entangled unitary synthesis. Finally, in Section~\ref{sec:conclusion}, we summarize our results and discuss potential future research directions.

\section{Preliminaries}
\label{sec:pre}

The core idea of utilizing measurements and feedback to accelerate quantum computation can be intuitively understood as follows. In our approach, we first identify the parts of the quantum circuit that typically require significant depth and transform them into operations centered around "copying" information. By leveraging measurements and feedback, the task of "copying" quantum information can be effectively reduced to classical computation. This enables us to replace a substantial portion of quantum computation with classical computation while preserving the correctness of the overall computation. Consequently, the circuit depth is significantly reduced.
Specifically, the act of "copying" quantum information is realized through the quantum fan-out gate \( F_n \), defined as:
\[
F_n \ket{x_0} \ket{x_1, \dots, x_n} = \ket{x_0} \ket{x_1 \oplus x_0, \dots, x_n \oplus x_0},
\]
where \( x_0, x_1, \dots, x_n \in \{0, 1\} \). Our result regarding \( F_n \) is presented in Theorem~\ref{the:mfanout}, where we introduce a novel method to reduce the number of ancilla. A similar result is achieved for GHZ state preparation, as stated in Theorem~\ref{the:mGHZ}.
The key innovation of our work, compared to previous studies \cite{baumer2024measurement,baumer2024efficient}, lies in our ability to reduce the number of ancilla qubits. Our approach involves combining two smaller quantum fan-out gates (or GHZ states) into a larger one using only two (or one) ancilla qubits. Detailed proofs and the extension to qudit systems for Theorem~\ref{the:mfanout} and Theorem~\ref{the:mGHZ} are provided in Appendix~\ref{app:GHZ}.

\begin{theorem}
\label{the:mfanout}
    The quantum fan-out gate \( F_n \) can be implemented using a constant-depth quantum circuit with \( n/c \) ancilla, where \( c \in \mathbb{Z}^{+} \) is a constant. This circuit requires only one layer of intermediate measurements.
\end{theorem}

\begin{theorem}
\label{the:mGHZ}
    The \( n \)-qubit GHZ state \( \frac{1}{\sqrt{2}} \ket{0}^{\otimes n} + \frac{1}{\sqrt{2}} \ket{1}^{\otimes n} \) can be prepared using a constant-depth quantum circuit. This circuit requires only one layer of intermediate measurements and \( n/c \) additional ancilla qubits, where \( c \in \mathbb{Z}^{+} \) is a constant.
\end{theorem}

The \( n \)-Toffoli gate (or \( n \)-controlled \( X \) gate) is defined as:
\begin{equation*}
    \ket{x_1, x_2, \dots, x_n}\ket{t} \to \ket{x_1, x_2, \dots, x_n}\ket{t \oplus \prod_{j=1}^{n} x_j}.
\end{equation*}
Here, we restate Theorem~\ref{the:mtoffoli} from \cite{takahashi2016collapse}. Additionally, as demonstrated in \cite{takahashi2016collapse}, the circuit size of \( O(n \log n) \) can be reduced to \( O(n \log^{(c)} n) \) for any positive constant \( c \), while maintaining a constant circuit depth. By applying this technique, we eliminate the factor of \( n \) in the circuit sizes \( O(n2^n) \) and \( O(n4^n) \) in Table~\ref{tab:result}. Notably, since the circuit depth is constant, the circuit size is bounded by \( O(n + m) \) when the circuit utilizes \( m \) ancilla.

\begin{theorem}
\label{the:mtoffoli}
    The \( n \)-Toffoli gate can be implemented using a constant-depth quantum circuit with a circuit size of \( O(n \log n) \) and \( O(n \log n) \) ancilla. This circuit comprises \( 6 \) layers of intermediate measurements \cite{takahashi2016collapse}.
\end{theorem}

In the following sections, the \( R_y(\theta) \) and \( Z(\theta) \) gates are utilized:

\begin{definition}
    Define $Ry(\theta)$ gate as follows \cite{nielsen2010quantum}:
    \begin{equation*}
        Ry(\theta) = \begin{pmatrix}
            \cos \frac{\theta}{2} & -\sin \frac{\theta}{2} \\
            \sin \frac{\theta}{2} & \cos \frac{\theta}{2}
        \end{pmatrix}.
    \end{equation*}
\end{definition}

\begin{definition}
    Define $Z(\theta)$ gate as follows:
    \begin{equation*}
        Z(\theta) = \begin{pmatrix}
            1 & 0 \\
            0 & e^{i\theta}
        \end{pmatrix}.
    \end{equation*}
\end{definition}

\section{Constant-depth circuit for quantum state preparation}
\label{sec:qsp}

Our goal is to construct a circuit that prepares the following \( n \)-qubit state:
\begin{equation*}
    \sum_{j = 0}^{2^n-1} \alpha_{j} e^{i\theta_{j}} \ket{j}, \quad \alpha_{j} \in [0,1], \quad \theta_{j} \in [0,2\pi), \quad \sum_{j=0}^{2^n-1} \alpha_j^2 = 1.
\end{equation*}

The construction is carried out in three steps. First, we prepare the state:
\[
\sum_{j=0}^{2^n-1} \alpha_{j} \ket{e_{j}},
\]
where \( e_j \) is the one-hot encoding of \( j \). Next, we introduce the phase \( e^{i\theta_j} \) to each basis state \( \ket{e_j} \). Finally, we transform each \( \ket{e_j} \) into \( \ket{j} \), thereby successfully preparing the desired state.

The most technically challenging aspect of this construction is addressed in Lemma~\ref{lem:monehot}:

\begin{lemma}
\label{lem:monehot}
    The following \( 2^n \)-qubit state can be prepared from the initial state \( \ket{0}^{\otimes 2^n} \) using a constant-depth quantum circuit with a size of \( O(n4^n) \). This circuit utilizes \( 22 \) layers of intermediate measurements and \( O(n4^n) \) ancilla.
    \begin{align*}
    &\sum_{j=0}^{2^n-1} \alpha_j \ket{e_{j}}, \ \ \ \ket{e_0} = \ket{1}\otimes \ket{0}^{2^n-1}, \mbox{and}\\ 
    &\ket{e_j} = \ket{0}^{\otimes j}\ket{1}\ket{0}^{\otimes (N-j-1)}, j=1,2,\dots,2^n-1.
\end{align*}
\end{lemma}

\begin{figure}[htbp]
    \centering
    \begin{tikzpicture}[
        level distance=2cm,  % 层级垂直间距
        level 1/.style={sibling distance=3cm},  % 第一层水平间距
        level 2/.style={sibling distance=3cm},  
        level 3/.style={sibling distance=3cm},
        node style/.style={
            draw=blue!50, 
            fill=blue!10, 
            rounded corners=3pt, 
            minimum width=1.8cm, 
            align=center, 
            font=\small\ttfamily,
            inner sep=3pt
        },
        edge style/.style={
            ->, 
            >=stealth, 
            thick, 
            blue!50,
            shorten >=2pt
        },
        control edge/.style={
            red!60!black,
            dashed,
            shorten >=4pt
        },
        gate label/.style={
            black,
            font=\footnotesize,
            midway,
            fill=white,
            inner sep=1pt
        }
    ]
    
    % 根节点：初始状态
    \node[node style] (Root) {$\ket{0000}$}
    % 第一层：应用第一个Ry门
    child {
        node[node style] (A) {$\ket{0000}$}
        % 第二层：应用第二个Ry门（控制条件：q0=0）
        child {
            node[node style] (B) {$\ket{0000}$}
            % 第三层：应用第三个Ry门（双控制）
            child {
                node[node style] (C) {$\ket{0000}$}
                % 第四层：应用三控制X门
                child {
                    node[node style, fill=red!20] (D) {$\ket{0001}$}
                    edge from parent[edge style]
                    node[gate label, pos=0.7] {$\ket{000}$-$X$}
                }
                edge from parent[edge style]
                node[gate label, pos=0.7] {$R_y(\theta_3)$}
            }
            child {
                node[node style, fill=red!20] (E) {$\ket{0010}$}
                edge from parent[edge style]
                node[gate label, pos=0.7] {$R_y(\theta_3)$}
            }
            edge from parent[edge style]
            node[gate label, pos=0.7] {$R_y(\theta_2)$}
        }
        child {
            node[node style, fill=red!20] (F) {$\ket{0100}$}
            edge from parent[edge style]
            node[gate label, pos=0.7] {$R_y(\theta_2)$}
        }
        edge from parent[edge style]
        node[gate label, pos=0.7] {$R_y(\theta_1)$}
    }
    child {
        node[node style, fill=red!20] (G) {$\ket{1000}$}
        edge from parent[edge style]
        node[gate label, pos=0.7] {$R_y(\theta_1)$}
    };

    % 添加控制线标注
    \draw[control edge] (A.south) -- ++(0,-0.2) -| node[pos=0.25, below] {\footnotesize Control: $q_0=0$} (F.north);
    \draw[control edge] (B.south) -- ++(0,-0.2) -| node[pos=0.25, below] {\footnotesize Controls: $q_0=q_1=0$} (E.north);

    % 添加最终状态标注
    \end{tikzpicture}
    \caption{The figure illustrates the direct method for quantum state preparation. In this approach, each qubit is processed sequentially.}
    \label{fig:mQSP1}
\end{figure}
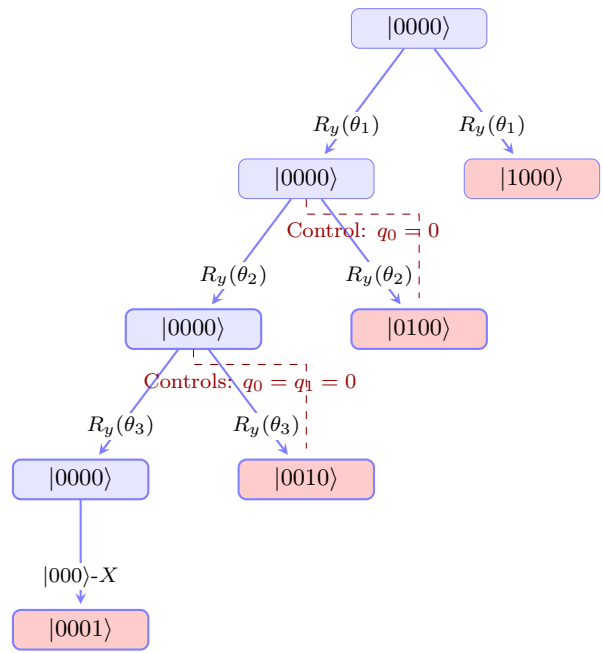

We will provide a high-level overview of the construction process. To illustrate the state preparation, we use the following example:
\begin{equation*}
    \alpha_1 \ket{1000} + \alpha_2 \ket{0100} + \alpha_3 \ket{0010} + \alpha_4 \ket{0001}.
\end{equation*}
A straightforward approach to preparing this state is as follows. 
First, apply an \( R_y(\theta_1) \) gate to the first qubit of the initial state \( \ket{0000} \), creating a superposition of \( \ket{0000} \) and \( \ket{1000} \). Next, apply an \( R_y(\theta_2) \) gate to the second qubit, conditioned on the first qubit being in the state \( \ket{0} \). This generates a superposition of \( \ket{0000} \), \( \ket{1000} \), and \( \ket{0100} \). Similarly, by applying a two-control \( R_y(\theta_3) \) gate, we extend the superposition to include \( \ket{0010} \). By carefully choosing the parameters \( \theta_1, \theta_2, \) and \( \theta_3 \), the resulting state becomes:
\[
\alpha_1 \ket{1000} + \alpha_2 \ket{0100} + \alpha_3 \ket{0010} + \alpha_4 \ket{0000}.
\]
Finally, we transform \( \ket{0000} \) into \( \ket{0001} \) by applying a three-control \( X \) gate, thereby completing the state preparation.
The conceptual diagram of this state preparation method is shown in Fig.~\ref{fig:mQSP1}. It is clear that the circuit depth scales at least linearly with the number of qubits.

\begin{figure*}[htbp]
    \centering
    \begin{tikzpicture}[
        level distance=1cm,
        level 1/.style={sibling distance=8cm},
        level 2/.style={sibling distance=4cm},
        level 3/.style={sibling distance=2cm},
        level 4/.style={sibling distance=2cm},
        node style/.style={
            draw=blue!50,
            fill=blue!10,
            rounded corners=3pt,
            minimum width=1.2cm,
            align=center,
            font=\small\ttfamily,
            inner sep=4pt
        },
        merge style/.style={
            draw=green!50,
            fill=green!10,
            double,
            double distance=1pt
        },
        final style/.style={
            draw=red!50,
            fill=red!20
        },
        edge style/.style={
            ->,
            >=stealth,
            thick,
            blue!50
        },
        merge edge/.style={
            ->,
            >=stealth,
            thick,
            green!50,
            dashed
        }
    ]
    
    % 第一层：初始扩展
    \node[node style] (Root) {$\ket{0000}$}
    child {
        node[node style] (L1) {$\ket{0000}$}
        child {
            node[node style] (L2a) {$\ket{0000}$}
            child {
                node[node style] (L3a) {$\ket{0000}$}
                child {
                node[final style] (L4) {$\ket{0001}$}
                edge from parent[edge style]
                }
                edge from parent[edge style]
            }
            child {
                node[final style] (L3b) {$\ket{0010}$}
                edge from parent[edge style]
            }
            edge from parent[edge style]
        }
        child {
            node[final style] (L2b) {$\ket{0100}$}
            child {
                node[merge style] (L3a) {$\ket{0100}$}
                edge from parent[merge edge]
            }
            child {
                node[merge style] (L3b) {$\ket{0110}$}
                edge from parent[merge edge]
            }
            edge from parent[edge style]
        }
        edge from parent[edge style]
    }
    child {
        node[final style] (R1) {$\ket{1000}$}
        child {
            node[merge style] (R2a) {$\ket{1000}$}
            child {
                node[merge style] (R3a) {$\ket{1000}$}
                edge from parent[merge edge]
            }
            child {
                node[merge style] (R3b) {$\ket{1010}$}
                edge from parent[merge edge]
            }
            edge from parent[merge edge]
        }
        child {
            node[merge style] (R2b) {$\ket{1100}$}
            child {
                node[merge style] (R3c) {$\ket{1100}$}
                edge from parent[merge edge]
            }
            child {
                node[merge style] (R3d) {$\ket{1110}$}
                edge from parent[merge edge]
            }
            edge from parent[merge edge]
        }
        edge from parent[edge style]
    };

    \end{tikzpicture}
    \caption{The figure illustrates the core concept of our quantum state preparation method. The process is divided into three key stages:
(1) expanding the superposition to include all relevant basis states,
(2) pruning unnecessary quantum states (indicated by dashed arrows) through parallel operations, and
(3) transforming \( \ket{0000} \) into \( \ket{0001} \) by applying a three-control \( X \) gate.}
    \label{fig:mQSP2}
\end{figure*}
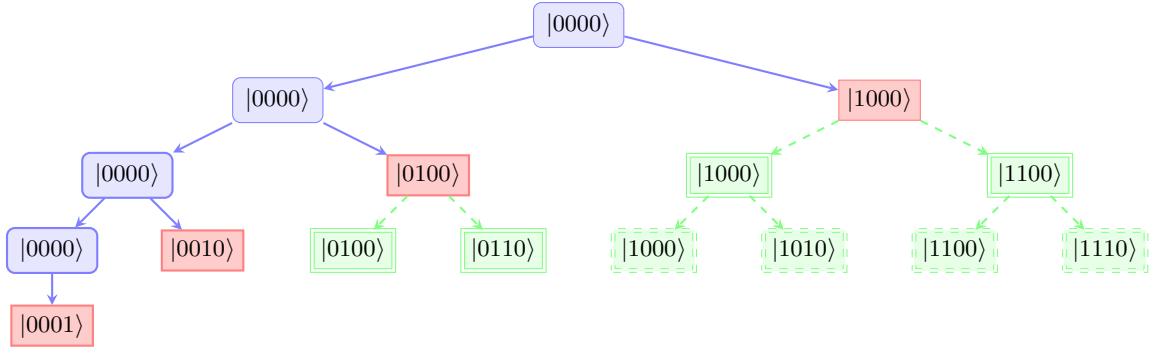

To avoid high circuit depth, we propose a novel method. First, we apply \( R_y(\theta_1) \), \( R_y(\theta_2) \), and \( R_y(\theta_3) \) gates to the first three qubits, generating a superposition of \( 8 \) basis states. Next, we combine specific basis states to simplify the superposition. For instance, when the first two qubits are in the state \( \ket{01} \), we combine the superposition of \( \ket{0100} \) and \( \ket{0110} \) into \( \ket{0100} \) by applying an \( R_y(-\theta_3) \) gate to the third qubit. Similarly, when the first qubit is in the state \( \ket{1} \), the superposition of \( \ket{1000} \), \( \ket{1010} \), \( \ket{1100} \), and \( \ket{1110} \) can be combined into \( \ket{1000} \) by applying \( R_y(-\theta_2) \) and \( R_y(-\theta_3) \) gates to the second and third qubits. Crucially, these combinations can be implemented in parallel. Finally, we transform \( \ket{0000} \) into \( \ket{0001} \) by applying a three-control \( X \) gate, thereby completing the state preparation.
The conceptual framework of our construction is illustrated in Fig.~\ref{fig:mQSP2}. The detailed proof is provided in Appendix~\ref{app:qsp}.

We now present Lemma~\ref{lem:mtobinary}, which transforms the quantum state from one-hot encoding (generated by Lemma~\ref{lem:monehot}) into its binary representation. While a similar result is proven in \cite{yeo2025reducing}, we provide a detailed proof in Appendix~\ref{app:qsp} for completeness.

\begin{lemma}
\label{lem:mtobinary}
    There exists a constant-depth circuit \( C \) with a size of \( O(n2^n \log n) \), utilizing \( O(n2^n \log n) \) ancilla qubits and \( 9 \) layers of intermediate measurements, such that for an arbitrary \( N \)-qubit state \( \sum_{j=0}^{N-1} \alpha_j \ket{e_j} \), where \( N = 2^n \), the following equation holds:
    \begin{equation*}
        C\sum_{j=0}^{N-1} \alpha_j \ket{e_j} = \sum_{j=0}^{N-1} \alpha_j \ket{j}\ket{0}^{\otimes (N-n)}.
    \end{equation*}
\end{lemma}

With all the necessary lemmas established, we now proceed to prove Theorem~\ref{the:mstatepre}. Importantly, the circuit size of \( O(n4^n) \) can be further optimized, as demonstrated in the following section.

\begin{theorem}
\label{the:mstatepre}
    An arbitrary \( n \)-qubit quantum state of the form \( \sum_{j=0}^{2^n-1} \alpha_j e^{i\theta_j} \ket{j} \) can be prepared from the initial state \( \ket{0}^{\otimes n} \) using a constant-depth quantum circuit. The circuit has a size of \( O(n4^n) \), utilizes \( O(n4^n) \) ancilla, and requires \( 31 \) layers of intermediate measurements.
\end{theorem}
\begin{proof}
    In Step 1, we prepare the state \( \sum_{j=0}^{2^n - 1} \alpha_{j} \ket{e_j} \) using Lemma~\ref{lem:monehot}.

In Step 2, for \( j = 0, 1, \dots, 2^n - 1 \), we apply a \( Z(\theta_j) \) gate to the \( (j+1) \)-th qubit, resulting in the state:
\[
\sum_{j=0}^{2^n - 1} \alpha_{j} e^{i\theta_j} \ket{e_j}.
\]

In Step 3, we transform the state into \( \sum_{j=0}^{2^n - 1} \alpha_{j} e^{i\theta_j} \ket{j} \) using Lemma~\ref{lem:mtobinary}.

The entire circuit has a constant depth. The circuit size and ancilla count are bounded by \( O(n4^n) \), and the circuit requires \( 31 \) layers of intermediate measurements, as specified in Lemma~\ref{lem:monehot} and Lemma~\ref{lem:mtobinary}.
\end{proof}

When the quantum state is sparse, the complexity of the circuit needed for its preparation can be substantially reduced, as demonstrated in Theorem~\ref{the:msparse}. A detailed proof of this result is provided in Appendix~\ref{app:qsp}.

\begin{theorem}
\label{the:msparse}
    For an arbitrary \( n \)-qubit sparse quantum state \( \vert\psi\rangle \) with \( s \) basis states having non-zero coefficients, there exists a constant-depth circuit \( C \) that utilizes \( O(s^2 \log n) \) ancilla qubits and \( 31 \) layers of intermediate measurements, such that:
    \begin{equation*}
        C \ket{0}^{\otimes n} =\ket{\psi}.
    \end{equation*}
\end{theorem}

\section{Constant-depth circuit for controlled quantum state preparation}
\label{sec:cqsp}

To extend quantum state preparation to a controlled version, we introduce a control qubit to each gate used in Theorem~\ref{the:mstatepre}. By initially "copying" the control qubit a sufficient number of times, we ensure that the overall circuit depth remains constant. This approach establishes Lemma~\ref{lem:mCQSP}, with the detailed proof provided in Appendix~\ref{app:cqsp}.

\begin{lemma}
\label{lem:mCQSP}
    For an arbitrary \( n \)-qubit state \( \sum_{j = 0}^{2^n - 1} \alpha_j \ket{j} \) and a one-qubit state \( a\ket{0} + b\ket{1} \), where \( a, b, \alpha_0, \dots, \alpha_{2^n - 1} \in \mathbb{C} \), a constant-depth quantum circuit \( C \) of size \( O(n4^n) \) can perform controlled state preparation. This circuit employs \( O(n4^n) \) ancilla and requires \( 33 \) layers of intermediate measurements.
    \begin{equation*}
        C(a \ket{0} + b \ket{1})\ket{0}^{\otimes n} = a\ket{0}^{\otimes (n+1)} + b \ket{1}\sum_{j=0}^{2^n-1} \alpha_j \ket{j}.
    \end{equation*}
\end{lemma}

When the quantum state preparation is controlled by \( n \) control qubits, the complexity of the problem increases significantly. At a high level, each quantum gate used in Theorem~\ref{the:mstatepre} should be replaced by a Uniformly Controlled Gate (UCG), defined as:
\begin{equation*}
    \ket{\boldsymbol{x}}\ket{t} \to \ket{\boldsymbol{x}}U_{\boldsymbol{x}}\ket{t}, \quad \boldsymbol{x} \in \{0,1\}^n.
\end{equation*}
Notably, UCGs can be implemented using a constant-depth circuit with the aid of measurements and feedback \cite{allcock2023constant}. Through careful circuit design, we establish Theorem~\ref{the:mcstatep}, with the detailed proof provided in Appendix~\ref{app:cqsp}.

\begin{theorem}
\label{the:mcstatep}
    Given arbitrary \( n \)-qubit states \( \ket{\psi_j} \) for \( j = 0, 1, \dots, 2^n - 1 \), we can construct a quantum circuit \( C \) such that for an arbitrary \( n \)-qubit state \( \sum_{j = 0}^{2^n - 1} \alpha_j \ket{j} \), where \( \alpha_1, \dots, \alpha_{2^n - 1} \in \mathbb{C} \), the following holds:
\begin{equation*}
    C \left( \sum_{j=0}^{2^n-1} \alpha_j \ket{j} \otimes \ket{0}^{\otimes n} \right) = \sum_{j=0}^{2^n-1} \alpha_j \ket{j} \otimes \ket{\psi_j}.
\end{equation*}
The circuit \( C \) has a constant depth, a size of \( O(n4^n) \), utilizes \( O(n4^n) \) ancilla, and requires \( 50 \) layers of intermediate measurements.
\end{theorem}

Entangled unitary synthesis can be regarded as a special case of quantum state preparation controlled by \( n \) qubits. As a direct result, we derive Corollary~\ref{cor:munitary}.

\begin{corollary}
\label{cor:munitary}
    Given an arbitrary \( 2^n \times 2^n \) unitary \( U \), we can construct a constant-depth quantum circuit \( C \) with a size of \( O(n4^n) \). This circuit utilizes \( O(n4^n) \) ancilla and requires \( 50 \) layers of intermediate measurements. For an arbitrary \( n \)-qubit state \( \sum_{j = 0}^{2^n - 1} \alpha_j \ket{j} \), where \( \alpha_1, \dots, \alpha_{2^n - 1} \in \mathbb{C} \), the following holds:
    \begin{equation*}
        C\sum_{j=0}^{2^n-1}\alpha_j \ket{j}\otimes \ket{0}^{\otimes n} = \sum_{j=0}^{2^n-1}\alpha_j \ket{j}\otimes U\ket{j}.
    \end{equation*}
\end{corollary}
\begin{proof}
    Note that for \( j = 0, 1, \dots, 2^n - 1 \), \( U\ket{j} \) is an \( n \)-qubit quantum state. Therefore, this corollary follows directly from Theorem~\ref{the:mcstatep}.
\end{proof}

By leveraging Theorem~\ref{the:mstatepre} and Theorem~\ref{the:mcstatep}, the circuit size for quantum state preparation can be reduced from \( O(n4^n) \) to \( O(n2^n) \), as demonstrated in Corollary~\ref{cor:mstatepre}.

\begin{corollary}
\label{cor:mstatepre}
    An arbitrary \( n \)-qubit quantum state can be prepared from the initial state \( \ket{0}^{\otimes n} \) using a constant-depth quantum circuit with a size of \( O(n2^n) \). This circuit utilizes \( O(n2^n) \) ancilla and requires \( 81 \) layers of intermediate measurements.
\end{corollary}
\begin{proof}
    For an arbitrary \( n \)-qubit state \( \sum_{j=0}^{2^n-1} \alpha_j \ket{j} \), where \( \alpha_0, \dots, \alpha_{2^n-1} \in \mathbb{R} \) and \( \sum_{j=0}^{2^n-1} |\alpha_j|^2 = 1 \), Theorem~\ref{the:mstatepre} guarantees that this state can be prepared using a constant-depth circuit with \( O(n4^n) \) ancilla.

For an arbitrary \( n \)-qubit state \( \sum_{k=0}^{2^n-1} \beta_{j,k} \ket{k} \), where \( j = 0, 1, \dots, 2^n - 1 \) and \( \beta_{j,k} \in \mathbb{C} \), Theorem~\ref{the:mcstatep} ensures the existence of a constant-depth circuit \( C \) with \( O(n4^n) \) ancilla that satisfies:
\begin{equation*}
    C \left( \sum_{j=0}^{2^n-1} \alpha_j \ket{j} \otimes \ket{0}^{\otimes n} \right) = \sum_{j=0}^{2^n-1} \alpha_j \ket{j} \otimes \sum_{k=0}^{2^n-1} \beta_{j,k} \ket{k}.
\end{equation*}

Notably, any \( 2n \)-qubit state can be expressed in the form \( \sum_{j=0}^{2^n-1} \alpha_j \ket{j} \otimes \sum_{k=0}^{2^n-1} \beta_{j,k} \ket{k} \). Therefore, an arbitrary \( 2n \)-qubit state can be prepared using a constant-depth circuit with \( O(n4^n) \) ancilla. Similarly, for an arbitrary \( (2n - 1) \)-qubit state \( \ket{\phi} \), the state \( \ket{\phi} \otimes \ket{0} \) can be prepared using a constant-depth circuit with \( O(n4^n) \) ancilla. Consequently, an arbitrary \( n \)-qubit state can be prepared using a constant-depth circuit with \( O(n2^n) \) ancilla. Since the circuit depth is constant, the circuit size is bounded by the number of qubits \( O(n2^n) \). This circuit requires \( 81 \) layers of intermediate measurements, as established by Theorem~\ref{the:mstatepre} and Theorem~\ref{the:mcstatep}.
\end{proof}

According to \cite{takahashi2016collapse}, in Theorem~\ref{the:mtoffoli}, the circuit size of \( O(n \log n) \) can be reduced to \( O(n \log^{(c)} n) \) for any positive constant \( c \), while maintaining a constant circuit depth. Here, \( \log^{(c)} n \) denotes the \( c \)-times iterated logarithm. In practice, by setting \( c = 5 \), the \( n \)-Toffoli gate can be implemented using a constant-depth circuit with a size of \( O(n) \). This is because, for all practical values of \( n \), \( \log^{(5)} n < 10 \). By applying this technique, we eliminate the factor of \( n \) in the circuit sizes \( O(n2^n) \) and \( O(n4^n) \), reducing them to \( O(2^n) \) and \( O(4^n) \), respectively, as shown in Table~\ref{tab:result}. A more detailed discussion is provided in Appendix~\ref{app:cqsp}.

As an additional result, we demonstrate that the reversible functions defined below can be implemented using a constant-depth quantum circuit with the aid of measurements and feedback, as stated in Theorem~\ref{the:refunction}. The detailed proof is provided in Appendix~\ref{app:refunction}.

\begin{definition}
    A function \( f: \{0,1\}^n \to \{0,1\}^n \) is reversible if and only if for any \( \boldsymbol{x}, \boldsymbol{y} \in \{0,1\}^n \) where \( \boldsymbol{x} \neq \boldsymbol{y} \), it holds that \( f(\boldsymbol{x}) \neq f(\boldsymbol{y}) \).
\end{definition}

\begin{theorem}
\label{the:refunction}
    For an arbitrary reversible function \( f: \{0,1\}^n \to \{0,1\}^n \), there exists a constant-depth circuit \( C \) with \( O(n2^n \log n) \) ancilla that implements the function as follows. This circuit requires \( 18 \) layers of intermediate measurements.
    \begin{equation*}
        C \ket{\boldsymbol{x}} = \ket{f(\boldsymbol{x})}, \quad \boldsymbol{x} \in \{0,1\}^n.
    \end{equation*}
\end{theorem}

\section{Conclusion and discussion}
\label{sec:conclusion}

By leveraging measurements and feedback, we have uncovered several novel results. We comprehensively demonstrate that quantum state preparation, controlled quantum state preparation, and entangled unitary synthesis can be achieved using constant-depth circuits. Furthermore, we conjecture that the circuit size is asymptotically optimal. Additionally, we show that any reversible function can be implemented using a constant-depth circuit. We also demonstrate that the GHZ state and the quantum fan-out gate can be implemented using constant-depth circuits with fewer ancilla, and we extend this result to qudit systems.

For future research directions, several intriguing questions arise. First, we aim to explore whether our current results can be further improved. Specifically, for unitary synthesis, we pose the following questions: What types of unitaries can be implemented using constant-depth circuits? Additionally, we consider the impact of qubit connectivity constraints. In particular, if qubit connectivity is restricted to a two-dimensional grid, can we still achieve the same results? Can these results be extended to qudit systems? Finally, we investigate the potential of using measurements and feedback to optimize general quantum circuits. Specifically, given an arbitrary quantum circuit, can its depth be reduced through the application of measurements and feedback?

% \section{acknowledgements}
% We thank Shuai Yang and Xiaoyu He for their insightful discussions.
%This work was supported in part by the National Natural Science Foundation of China under Grants No. 62325210 and 92465202, as well as the Strategic Priority Research Program of the Chinese Academy of Sciences under Grant No. XDB28000000.

\appendix
\section{Constant depth circuit for quantum fan-out}
\label{app:GHZ}

\begin{definition}
    Define the quantum fan-out gate $F_n$ as follows:
    \begin{equation*}
        F_n\ket{x_0}\ket{x_1,\cdots,x_n} = \ket{x_0}\ket{x_1 \oplus x_0, \cdots, x_n \oplus x_0}.
    \end{equation*}
\end{definition}

\begin{lemma}
\label{lem:GHZ}
    Given a qubit in the state \( \ket{\psi} = \alpha \ket{0} + \beta \ket{1} \), the quantum state \( \alpha \ket{0}^{\otimes n} + \beta \ket{1}^{\otimes n} \) can be prepared using a constant-depth quantum circuit. This circuit requires one layer of intermediate measurements and \( n/c \) additional ancilla qubits, where \( c \in \mathbb{Z}^{+} \) is a constant.
\end{lemma}

\begin{proof}
    \begin{figure*}[htbp]
        \centering
        \begin{minipage}{0.9\textwidth}
        \Qcircuit @C=0.85em @R=1em @!R{
        \lstick{\ket{\psi}} & \ctrl{1} & \qw &&& \ctrl{1} & \ctrl{2} & \ctrl{3} & \ctrl{4} & \qw &&& \ctrl{1} & \qw & \qw & \qw & \qw &&& \ctrl{1} & \qw & \qw & \qw \\
        \lstick{\ket{0}} & \targ \qwx[1] & \qw &&& \targ & \qw & \qw & \qw & \qw &&& \targ & \ctrl{1} & \qw & \qw & \qw &&& \targ & \ctrl{2} & \ctrl{1} & \qw \\
        \lstick{\ket{0}} & \targ \qwx[1] & \qw &\push{OR}&& \qw & \targ & \qw & \qw & \qw &\push{OR}&& \qw & \targ & \ctrl{1} & \qw & \qw &\push{OR}&& \qw & \qw & \targ & \qw \\
        \lstick{\ket{0}} & \targ \qwx[1] & \qw &&& \qw & \qw & \targ & \qw & \qw &&& \qw & \qw & \targ & \ctrl{1} & \qw &&& \qw & \targ & \ctrl{1} & \qw \\
        \lstick{\ket{0}} & \targ & \qw &&& \qw & \qw & \qw & \targ & \qw &&& \qw & \qw & \qw & \targ & \qw &&& \qw & \qw & \targ & \qw \\
        }
        \end{minipage}
        \caption{Given qubit in state $\ket{\psi} = \alpha \ket{0} + \beta \ket{1}$, the circuits that prepared $\alpha \ket{0}^{\otimes 5} + \beta \ket{1}^{\otimes 5}$.}
        \label{fig:fan-out}
    \end{figure*}
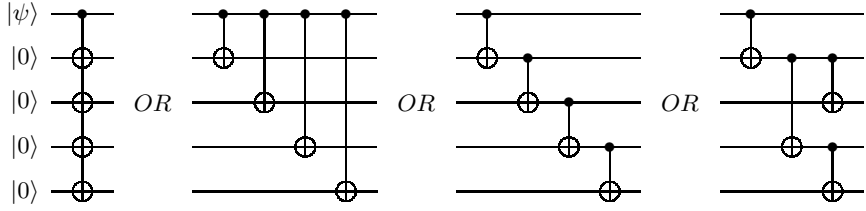
    For an arbitrary constant \( k \), the state \( \alpha \ket{0}^{\otimes k} + \beta \ket{1}^{\otimes k} \) can be prepared using a constant-depth circuit by employing well-known techniques. An example for the case where \( k = 5 \) is depicted in Fig.~\ref{fig:fan-out}. Similarly, the state \( \frac{1}{\sqrt{2}} \ket{0}^{\otimes k} + \frac{1}{\sqrt{2}} \ket{1}^{\otimes k} \) can also be prepared using a constant-depth circuit.

    In Step 1, the following state can be prepared within a constant-depth circuit:
    \begin{align*}
        &\left(\alpha \ket{0}^{\otimes k} + \beta \ket{1}^{\otimes k} \right) \bigotimes_{j=1}^m \left(\frac{1}{\sqrt{2}} \ket{0}^{\otimes k}  + \frac{1}{\sqrt{2}} \ket{1}^{\otimes k} \right) \\
        =& \frac{\alpha}{2^{m/2}} \ket{0}^{\otimes k} \sum_{a_1,\dots,a_m \in \{0,1\}} \bigotimes_{j=1}^m \ket{a_j}^{\otimes k} \\
        & + \frac{\beta}{2^{m/2}} \ket{1}^{\otimes k} \sum_{b_1,\dots,b_m \in \{0,1\}} \bigotimes_{j=1}^m \ket{b_j}^{\otimes k}.
    \end{align*}
    Note that there are \( m + 1 \) parts of qubits, each containing \( k \) qubits. The \( 0 \)-th part is in the state \( \alpha \ket{0}^{\otimes k} + \beta \ket{1}^{\otimes k} \), while the \( j \)-th part (for \( j = 1, 2, \dots, m \)) is in the state \( \frac{1}{\sqrt{2}} \ket{0}^{\otimes k} + \frac{1}{\sqrt{2}} \ket{1}^{\otimes k} \). Let \( \ket{a_0} = \ket{0} \) and \( \ket{b_0} = \ket{1} \).

    In Step 2, we introduce \( m \) ancilla. For \( j = 1, 2, \dots, m \), we compute the parity between the \( (j-1) \)-th part and the \( j \)-th part of qubits, storing the result in the \( j \)-th ancilla. This results in the following state:
    \begin{align*}
        &\frac{\alpha}{2^{m/2}} \ket{a_0}^{\otimes k} \sum_{a_1,\dots,a_m \in \{0,1\}} \bigotimes_{j=1}^m \ket{a_j}^{\otimes k}\ket{a_{j-1} \oplus a_j}\\
        +& \frac{\beta}{2^{m/2}} \ket{b_0}^{\otimes k} \sum_{b_1,\dots,b_m \in \{0,1\}} \bigotimes_{j=1}^m \ket{b_j}^{\otimes k}\ket{b_{j-1} \oplus b_j}.
    \end{align*}
    The state of the \( j \)-th ancilla is denoted by \( \ket{a_{j-1} \oplus a_j} \) and \( \ket{b_{j-1} \oplus b_j} \), as shown in the above formula. This state is achieved by applying two CNOT gates to the \( j \)-th ancilla. One CNOT gate is controlled by an arbitrary qubit in the \( (j-1) \)-th part, and the other is controlled by an arbitrary qubit in the \( j \)-th part.

    In Step 3, all ancilla are measured in the \( \{\ket{0}, \ket{1}\} \) basis, yielding measurement results \( c_1, \dots, c_m \in \{0,1\} \). The quantum state is then transformed as follows:
    \begin{align*}
        \alpha \ket{\hat{a}_0}^{\otimes k} \bigotimes_{j=1}^m \ket{\hat{a}_j}^{\otimes k}
        + \beta \ket{\hat{b}_0}^{\otimes k} \bigotimes_{j=1}^m \ket{\hat{b}_j}^{\otimes k}.
    \end{align*}
    Note that \( \ket{\hat{a}_0} = \ket{0} \) and \( \ket{\hat{b}_0} = \ket{1} \). For \( j = 1, 2, \dots, m \):
    \begin{align*}
        \bigoplus_{p=1}^j c_p &= \bigoplus_{p=1}^j \left( \hat{a}_p \oplus \hat{a}_{p-1} \right) = \hat{a}_0 \oplus \hat{a}_j \\
        &=\bigoplus_{p=1}^j \left( \hat{b}_p \oplus \hat{b}_{p-1} \right) = \hat{b}_0 \oplus \hat{b}_j.
    \end{align*}
    Thus, we have:
    \begin{align*}
        0 = \hat{a}_0 = \hat{a}_j \oplus \bigoplus_{p=1}^j c_p, \ \ \
        1 = \hat{b}_0 = \hat{b}_j \oplus \bigoplus_{p=1}^j c_p.
    \end{align*}
    
    In Step 4, for each \( j = 1, 2, \dots, m \), if \( \bigoplus_{p=1}^j c_p = 1 \), we apply the \( X \) gate to each qubit in the \( j \)-th part. This yields the desired state:
    \begin{equation*}
        \alpha \ket{0}^{\otimes k(m+1)} + \beta \ket{1}^{\otimes k(m+1)}.
    \end{equation*}

    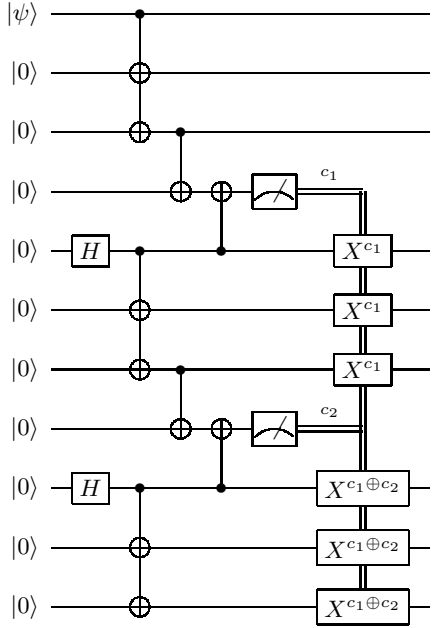
\begin{figure}
        \centering
        \begin{minipage}{0.9\textwidth}
        \Qcircuit @C=0.85em @R=1em @!R{
        \lstick{\ket{\psi}} & \qw & \ctrl{1} & \qw & \qw & \qw & \qw & \qw \\
        \lstick{\ket{0}} & \qw & \targ \qwx[1] & \qw & \qw & \qw & \qw & \qw \\
        \lstick{\ket{0}} & \qw & \targ & \ctrl{1} & \qw & \qw & \qw & \qw \\
        \lstick{\ket{0}} & \qw & \qw & \targ & \targ & \meter & \ustick{_{c_1} \ \ \ \ \ \ \ \ } \cw \cwx[1] &  \\
        \lstick{\ket{0}} & \gate{H} & \ctrl{1} & \qw & \ctrl{-1} & \qw & \gate{X^{c_1}} \cwx[1] & \qw \\
        \lstick{\ket{0}} & \qw & \targ \qwx[1] & \qw & \qw & \qw & \gate{X^{c_1}} \cwx[1] & \qw \\
        \lstick{\ket{0}} & \qw & \targ & \ctrl{1} & \qw & \qw & \gate{X^{c_1}} \cwx[1] & \qw \\
        \lstick{\ket{0}} & \qw & \qw & \targ & \targ & \meter & \ustick{_{c_2} \ \ \ \ \ \ \ \ } \cw \cwx[1] &  \\
        \lstick{\ket{0}} & \gate{H} & \ctrl{1} & \qw & \ctrl{-1} & \qw & \gate{X^{c_1 \oplus c_2}} \cwx[1] & \qw \\
        \lstick{\ket{0}} & \qw & \targ \qwx[1] & \qw & \qw & \qw & \gate{X^{c_1 \oplus c_2}} \cwx[1] & \qw \\
        \lstick{\ket{0}} & \qw & \targ & \qw & \qw & \qw & \gate{X^{c_1 \oplus c_2}} & \qw \\
        }
        \end{minipage}
        \caption{Given qubit in state $\ket{\psi} = \alpha \ket{0} + \beta \ket{1}$. An instance for preparing state $\alpha \ket{0}^{\otimes 9} + \beta \ket{1}^{\otimes 9}$ using two ancilla.}
        \label{fig:fan-out2}
    \end{figure}
    
    For \( k = 3 \) and \( m = 2 \), an example is illustrated in Fig.~\ref{fig:fan-out2}. The entire circuit has a constant depth, and only one layer of intermediate measurements is used in Step 3. By introducing \( m \) additional ancilla qubits, we can prepare an \( n = k(m + 1) \)-qubit state for an arbitrary constant \( k \). To satisfy the requirement, we can set \( k \geq \frac{nc}{n + c} \), thereby completing the proof.
\end{proof}

\begin{corollary}
    The \( n \)-qubit GHZ state \( \frac{1}{\sqrt{2}} \ket{0}^{\otimes n} + \frac{1}{\sqrt{2}} \ket{1}^{\otimes n} \) can be prepared using a constant-depth circuit. This circuit requires one layer of intermediate measurements and \( n/c \) additional ancilla, where \( c \in \mathbb{Z}^{+} \) is a constant.
\end{corollary}
\begin{proof}
    Initially, we start with the state \( \ket{0}^{\otimes n} \). We then apply an \( H \) gate to the first qubit, resulting in the state:
    \begin{equation*}
        \frac{1}{\sqrt{2}} (\ket{0} + \ket{1}) \ket{0}^{\otimes (n-1)}.
    \end{equation*}
    Using Lemma~\ref{lem:GHZ}, we obtain the desired state:
    \begin{equation*}
        \frac{1}{\sqrt{2}} \ket{0}^{\otimes n} + \frac{1}{\sqrt{2}} \ket{1}^{\otimes n}.
    \end{equation*}
    This completes the proof.
\end{proof}

\begin{lemma}
\label{lem:recover}
    Given \( n \) qubits in the state \( \alpha \ket{0}^{\otimes n} + \beta \ket{1}^{\otimes n} \), a constant-depth circuit can recover a single qubit in the state \( \alpha \ket{0} + \beta \ket{1} \). This circuit requires one layer of intermediate measurements and no ancilla. Additionally, the remaining \( n - 1 \) qubits are measured and then released.
\end{lemma}
\begin{proof}
    Initially, we have \( n \) qubits in the state:
    \begin{equation*}
        \alpha \ket{0}^{\otimes n} + \beta \ket{1}^{\otimes n}.
    \end{equation*}
    In Step 1, we apply \( H \) gates to the last \( n - 1 \) qubits, resulting in:
    \begin{equation*}
        \alpha \ket{0} \left( \frac{\ket{0} + \ket{1}}{\sqrt{2}} \right)^{\otimes (n-1)} + \beta \ket{1} \left( \frac{\ket{0} - \ket{1}}{\sqrt{2}} \right)^{\otimes (n-1)}.
    \end{equation*}
    In Step 2, we measure the last \( n - 1 \) qubits, obtaining measurement results \( c_1, c_2, \dots, c_{n-1} \in \{0,1\} \). The state then becomes:
    \begin{equation*}
        \alpha \ket{0} + (-1)^{\bigoplus_{j=1}^{n-1} c_j} \beta \ket{1}.
    \end{equation*}
    In Step 3, we classically compute \( \bigoplus_{j=1}^{n-1} c_j \). If the result is \( 1 \), we apply a \( Z \) gate to the remaining qubit. This yields the desired state:
    \begin{equation*}
        \alpha \ket{0} + \beta \ket{1}.
    \end{equation*}
    The remaining \( n - 1 \) qubits are measured and can be reset for reuse in subsequent circuits. The entire circuit has a constant depth, requires only one layer of intermediate measurements (in Step 2), and uses no ancilla.
\end{proof}

\begin{theorem}
\label{the:fanout}
    The quantum fan-out gate \( F_n \) can be implemented using a constant-depth quantum circuit with \( n/c \) ancilla, where \( c \in \mathbb{Z}^{+} \) is a constant. This circuit requires one layer of intermediate measurements.
\end{theorem}

\begin{proof}
    \begin{figure}
        \centering
        \begin{minipage}{0.9\textwidth}
        \Qcircuit @C=0.85em @R=1em @!R{
        \lstick{\ket{x_0}} & \qw & \qw & \ctrl{1} & \qw & \qw & \gate{Z^{c_2}} & \qw \\
        \lstick{\ket{0}_1} & \gate{H} & \ctrl{1} & \targ & \qw & \meter & \cw & \lstick{c_1}  \\
        \lstick{\ket{x_1}} & \qw & \targ \qwx[1] & \qw & \qw & \qw & \gate{X^{c_1}} & \qw \\
        \lstick{\ket{x_2}} & \qw & \targ \qwx[1] & \qw & \qw & \qw & \gate{X^{c_1}} & \qw \\
        \lstick{\ket{x_3}} & \qw & \targ \qwx[1] & \qw & \qw & \qw & \gate{X^{c_1}} & \qw \\
        \lstick{\ket{0}_2} & \qw & \targ & \ctrl{1} & \gate{H} & \meter & \cw & \lstick{c_2} \\
        \lstick{\ket{0}_3} & \gate{H} & \ctrl{1} & \targ & \qw & \meter & \cw & \lstick{c_3} \\
        \lstick{\ket{x_4}} & \qw & \targ \qwx[1] & \qw & \qw & \qw & \gate{X^{c_1\oplus c_3}} & \qw \\
        \lstick{\ket{x_5}} & \qw & \targ \qwx[1] & \qw & \qw & \qw & \gate{X^{c_1\oplus c_3}} & \qw \\
        \lstick{\ket{x_6}} & \qw & \targ & \qw & \qw & \qw & \gate{X^{c_1\oplus c_3}} & \qw \\
        }
        \end{minipage}
        \caption{Quantum circuit implementing the quantum fan-out gate \( F_6 \).}
        \label{fig:fan-out3}
    \end{figure}
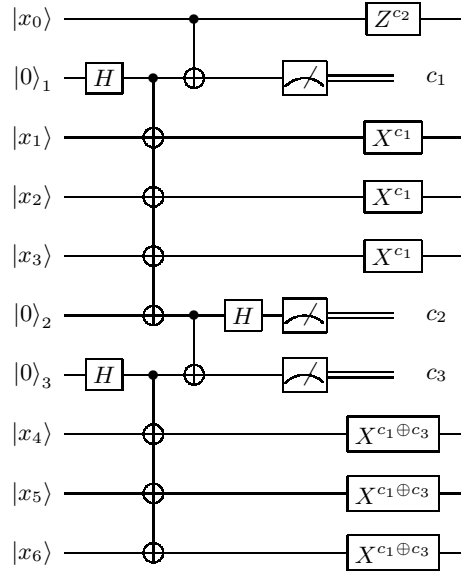

    This theorem generalizes the results presented in \cite{baumer2024efficient,baumer2024measurement}, incorporating a trade-off between ancilla count and circuit depth. Assume \( n = mp \), where \( m, p \in \mathbb{Z}^{+} \) and \( p \) is a constant. The quantum fan-out gate \( F_n \) performs the following transformation:
    \begin{equation*}
        \ket{x_0} \bigotimes_{j=1}^{m} \bigotimes_{k=1}^{p} \ket{x_{j,k}} \to \ket{x_0} \bigotimes_{j=1}^{m} \bigotimes_{k=1}^{p} \ket{x_{j,k} \oplus x_0}.
    \end{equation*}

    We use \( 2m - 1 \) ancilla, initialized to \( \ket{0} \), denoted as \( \ket{0}_1, \ket{0}_2, \dots, \ket{0}_{2m-1} \). The initial state, including ancilla, is:
    \begin{equation*}
        \ket{x_0} \left( \bigotimes_{j=1}^{m-1} \ket{0}_{2j-1} \left( \bigotimes_{k=1}^{p} \ket{x_{j,k}} \right) \ket{0}_{2j} \right) \ket{0}_{2m-1} \bigotimes_{k=1}^{p} \ket{x_{m,k}}.
    \end{equation*}

    In Step 1, apply \( H \) gates to the \( (2j-1) \)-th ancilla for \( j = 1, 2, \dots, m \), resulting in:
    \begin{align*}
        &\frac{1}{2^{m/2}} \sum_{a_1, \dots, a_m \in \{0,1\}} \ket{x_0} \left( \bigotimes_{j=1}^{m-1} \ket{a_j}_{2j-1} \left( \bigotimes_{k=1}^{p} \ket{x_{j,k}} \right) \ket{0}_{2j} \right) \\
        & \otimes \ket{a_m}_{2m-1} \bigotimes_{k=1}^{p} \ket{x_{m,k}}.
    \end{align*}

    In Step 2, use \( F_p \) and \( F_{p+1} \) gates to obtain:
    \begin{align*}
        &\frac{1}{2^{m/2}} \sum_{a_1, \dots, a_m \in \{0,1\}} \ket{x_0} \\
        & \otimes \left( \bigotimes_{j=1}^{m-1} \ket{a_j}_{2j-1} \left( \bigotimes_{k=1}^{p} \ket{x_{j,k} \oplus a_j} \right) \ket{a_j}_{2j} \right) \\
        & \otimes \ket{a_m}_{2m-1} \bigotimes_{k=1}^{p} \ket{x_{m,k} \oplus a_m}.
    \end{align*}
    Here, \( F_{p+1} \) gates add \( a_j \) to \( \ket{x_{j,1}}, \dots, \ket{x_{j,p}} \) and \( \ket{0}_{2j} \) for \( j = 1, 2, \dots, m-1 \), while one \( F_p \) gate adds \( a_m \) to \( \ket{x_{m,1}}, \dots, \ket{x_{m,p}} \). Since \( p \) is a constant, \( F_p \) and \( F_{p+1} \) gates can be implemented using constant-depth circuits, ensuring Step 2 has constant depth.

    In Step 3, apply a layer of CNOT gates to obtain:
    \begin{align*}
        &\frac{1}{2^{m/2}} \sum_{a_1, \dots, a_m \in \{0,1\}} \ket{x_0} \ket{a_1 \oplus x_0}_1 \left( \bigotimes_{k=1}^{p} \ket{x_{1,k} \oplus a_1} \right) \ket{a_1}_2 \\
        & \otimes \left( \bigotimes_{j=2}^{m-1} \ket{a_j \oplus a_{j-1}}_{2j-1} \left( \bigotimes_{k=1}^{p} \ket{x_{j,k} \oplus a_j} \right) \ket{a_j}_{2j} \right) \\
        & \otimes \ket{a_m \oplus a_{m-1}}_{2m-1} \bigotimes_{k=1}^{p} \ket{x_{m,k} \oplus a_m}.
    \end{align*}
    Here, \( x_0 \) is added to \( \ket{a_1} \), and \( a_j \) is added to \( \ket{a_{j+1}} \) for \( j = 1, 2, \dots, m-1 \). Steps 4 and 5 can be executed simultaneously but are described separately for clarity.

    In Step 4, measure the \( (2j-1) \)-th ancilla for \( j = 1, 2, \dots, m \), obtaining measurement results \( c_{2j-1} \). The remaining quantum state is:
    \begin{equation*}
        \ket{x_0} \left( \bigotimes_{j=1}^{m-1} \left( \bigotimes_{k=1}^{p} \ket{x_{j,k} \oplus \hat{a}_j} \right) \ket{\hat{a}_j}_{2j} \right) \bigotimes_{k=1}^{p} \ket{x_{m,k} \oplus \hat{a}_m}.
    \end{equation*}
    The following relations hold:
    \begin{align*}
        & c_1 = \hat{a}_1 \oplus x_0, \\
        & c_{2j-1} = \hat{a}_j \oplus \hat{a}_{j-1}, \quad j = 2, 3, \dots, m, \\
        & \bigoplus_{k=1}^{j} c_{2k-1} = \hat{a}_j \oplus x_0.
    \end{align*}
    For \( j = 1, 2, \dots, m \), classically compute \( \bigoplus_{k=1}^{j} c_{2k-1} \). If the result is \( 1 \), apply \( X \) gates to each qubit in \( \bigotimes_{k=1}^{p} \ket{x_{j,k} \oplus \hat{a}_j} \), yielding:
    \begin{equation*}
        \ket{x_0} \left( \bigotimes_{j=1}^{m-1} \left( \bigotimes_{k=1}^{p} \ket{x_{j,k} \oplus x_0} \right) \ket{\hat{a}_j}_{2j} \right) \bigotimes_{k=1}^{p} \ket{x_{m,k} \oplus x_0}.
    \end{equation*}

    In Step 5, apply \( H \) gates to the \( 2j \)-th ancilla for \( j = 1, 2, \dots, m-1 \), then measure these ancilla to obtain results \( c_{2j} \). The resulting quantum state is:
    \begin{equation*}
        (-1)^{\bigoplus_{j=1}^{m-1} \hat{a}_j c_{2j}} \ket{x_0} \bigotimes_{j=1}^{m} \bigotimes_{k=1}^{p} \ket{x_{j,k} \oplus x_0}.
    \end{equation*}
    Classically compute \( \bigoplus_{j=1}^{m-1} c_{2j} \). If the result is \( 1 \), apply a \( Z \) gate to \( \ket{x_0} \), resulting in:
    \begin{equation*}
        (-1)^{\bigoplus_{j=1}^{m-1} (\hat{a}_j c_{2j} \oplus x_0 c_{2j})} \ket{x_0} \bigotimes_{j=1}^{m} \bigotimes_{k=1}^{p} \ket{x_{j,k} \oplus x_0}.
    \end{equation*}
    Since \( x_0 \oplus \hat{a}_j = \bigoplus_{k=1}^{j} c_{2k-1} \) is a constant, the phase term \( (-1)^{\bigoplus_{j=1}^{m-1} (\hat{a}_j c_{2j} \oplus x_0 c_{2j})} \) is a global phase and can be omitted. Thus, we obtain the desired state:
    \begin{equation*}
        \ket{x_0} \bigotimes_{j=1}^{m} \bigotimes_{k=1}^{p} \ket{x_{j,k} \oplus x_0}.
    \end{equation*}

    An example circuit implementing \( F_6 \) with \( m = 2 \) and \( p = 3 \) is shown in Fig.~\ref{fig:fan-out3}. The entire circuit has constant depth, and one layer of intermediate measurements suffices for Steps 4 and 5. To use fewer than \( n/c \) ancilla, set \( m \leq \frac{n + c}{2c} \).
\end{proof}

The Parity gate \( P_n \) computes the parity of \( n \) qubits and stores the result in a target qubit. It is defined by the following transformation:
\begin{equation*}
    \ket{x_1, \dots, x_n} \ket{y} \to \ket{x_1, \dots, x_n} \ket{y \oplus \bigoplus_{j=1}^{n} x_j}.
\end{equation*}

According to \cite{hoyer2005quantum}, the Parity gate and the quantum fan-out gate are equivalent when constant-depth single-qubit gates are ignored. This directly leads to the following corollary:

\begin{corollary}
\label{cor:parity}
    The Parity gate \( P_n \) can be implemented using a constant-depth quantum circuit with \( n/c \) ancilla, where \( c \in \mathbb{Z}^{+} \) is a constant. This circuit requires one layer of intermediate measurements.
\end{corollary}

\subsection{The Quantum Fan-Out Gate in Qudit Systems}

A \( d \)-level qudit is associated with a \( d \)-dimensional Hilbert space, where \( d \geq 3 \) and \( d \) is an integer \cite{wang2020qudits,zi2023optimal}. Let \( \ket{0}, \ket{1}, \dots, \ket{d-1} \) denote the computational basis of a \( d \)-level qudit. Any state of a \( d \)-level qudit can be expressed as \( \ket{\psi} = \sum_{j=0}^{d-1} \alpha_j \ket{j} \), where each \( \alpha_j \in \mathbb{C} \) is a complex number satisfying \( \sum_{j=0}^{d-1} |\alpha_j|^2 = 1 \). Throughout this paper, \( d \) is treated as a constant.

\begin{definition}
    The quantum fan-out gate \( F_{n,d} \) in a \( d \)-dimensional qudit system is defined as:
    \begin{equation*}
        F_{n,d} \ket{x_0} \ket{x_1, \dots, x_n} = \ket{x_0} \ket{x_1 +_d x_0, \dots, x_n +_d x_0},
    \end{equation*}
    where \( +_d \) denotes addition modulo \( d \), and \( x_0, x_1, \dots, x_n \in \{0, 1, \dots, d-1\} \).
\end{definition}

\begin{definition}
    The \( C \)-\( X_d \) gate in a \( d \)-dimensional qudit system is defined as:
    \begin{equation*}
        C\text{-}X_d \ket{x} \ket{y} = \ket{x} \ket{y +_d x}, \quad x, y \in \{0, 1, \dots, d-1\}.
    \end{equation*}
\end{definition}

\begin{definition}
    The \( C \)-\( X^-_d \) gate in a \( d \)-dimensional qudit system is defined as:
    \begin{equation*}
        C\text{-}X^-_d \ket{x} \ket{y} = \ket{x} \ket{y -_d x}, \quad x, y \in \{0, 1, \dots, d-1\},
    \end{equation*}
    where \( -_d \) denotes subtraction modulo \( d \).
\end{definition}

\begin{definition}
    The \( H_d \) gate in a \( d \)-dimensional qudit system is defined as:
    \begin{equation*}
        H_d \ket{x} = \frac{1}{\sqrt{d}} \sum_{y=0}^{d-1} w_d^{xy} \ket{y}, x \in \{0, \dots, d-1\}, w_d = e^{2i\pi / d}.
    \end{equation*}
\end{definition}

\begin{definition}
    The \( X_{+c} \) gate in a \( d \)-dimensional qudit system is defined as:
    \begin{equation*}
        X_{+c} \ket{x} = \ket{x +_d c}, \quad x \in \{0, 1, \dots, d-1\}.
    \end{equation*}
\end{definition}

\begin{definition}
    The \( Z_d \) gate in a \( d \)-dimensional qudit system is defined as:
    \begin{equation*}
        Z_d \ket{x} = w_d^x \ket{x}, \quad x \in \{0, 1, \dots, d-1\}, \quad w_d = e^{2i\pi / d}.
    \end{equation*}
\end{definition}

\begin{lemma}
\label{lem:GHZ_d}
    Given a \( d \)-dimensional qudit in the state \( \ket{\psi} = \sum_{j=0}^{d-1} \alpha_j \ket{j} \), where \( \alpha_j \in \mathbb{C} \) and \( \sum_{j=0}^{d-1} |\alpha_j|^2 = 1 \), the quantum state \( \sum_{j=0}^{d-1} \alpha_j \ket{j}^{\otimes n} \) can be prepared using a constant-depth quantum circuit with \( n/c \) additional ancillary qudits. This circuit requires one layer of intermediate measurements, where \( c \in \mathbb{Z}^{+} \) is a constant.
\end{lemma}

\begin{proof}
    \begin{figure*}
        \centering
        \begin{minipage}{0.9\textwidth}
        \Qcircuit @C=0.85em @R=1em @!R{
        \lstick{\ket{\psi}} & \ctrl{1} & \qw &&& \ctrl{1} & \ctrl{2} & \ctrl{3} & \ctrl{4} & \qw &&& \ctrl{1} & \qw & \qw & \qw & \qw &&& \ctrl{1} & \qw & \qw & \qw \\
        \lstick{\ket{0}} & \gate{X_d} \qwx[1] & \qw &&& \gate{X_d} & \qw & \qw & \qw & \qw &&& \gate{X_d} & \ctrl{1} & \qw & \qw & \qw &&& \gate{X_d} & \ctrl{2} & \ctrl{1} & \qw \\
        \lstick{\ket{0}} & \gate{X_d} \qwx[1] & \qw &\push{OR}&& \qw & \gate{X_d} & \qw & \qw & \qw &\push{OR}&& \qw & \gate{X_d} & \ctrl{1} & \qw & \qw &\push{OR}&& \qw & \qw & \gate{X_d} & \qw \\
        \lstick{\ket{0}} & \gate{X_d} \qwx[1] & \qw &&& \qw & \qw & \gate{X_d} & \qw & \qw &&& \qw & \qw & \gate{X_d} & \ctrl{1} & \qw &&& \qw & \gate{X_d} & \ctrl{1} & \qw \\
        \lstick{\ket{0}} & \gate{X_d} & \qw &&& \qw & \qw & \qw & \gate{X_d} & \qw &&& \qw & \qw & \qw & \gate{X_d} & \qw &&& \qw & \qw & \gate{X_d} & \qw \\
        }
        \end{minipage}
        \caption{Given a qudit in the state \( \ket{\psi} = \sum_{j=0}^{d-1} \alpha_j \ket{j} \), the circuit prepares the state \( \sum_{j=0}^{d-1} \alpha_j \ket{j}^{\otimes 5} \). The leftmost circuit uses one \( F_{4,d} \) gate, while the other circuits consist of \( C \)-\( X_d \) gates.}
        \label{fig:fan-out_d}
    \end{figure*}
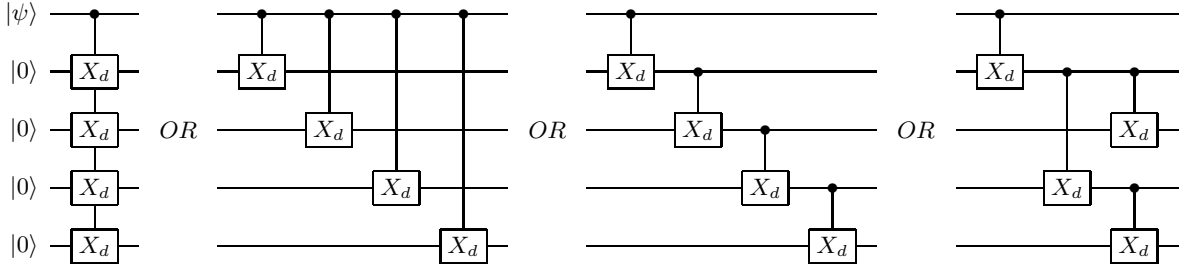

    For an arbitrary constant \( k \), the state \( \sum_{j=0}^{d-1} \alpha_j \ket{j}^{\otimes k} \) can be prepared using a constant-depth circuit by employing well-known methods. An example for \( k = 5 \) is shown in Fig.~\ref{fig:fan-out_d}. Similarly, the state \( \sum_{j=0}^{d-1} \frac{1}{\sqrt{d}} \ket{j}^{\otimes k} \) can be prepared by first applying an \( H_d \) gate to the first qudit at the beginning of the circuit.

    In Step 1, the following state can be prepared using a constant-depth circuit:
    \begin{align*}
        &\left( \sum_{j=0}^{d-1} \alpha_j \ket{j}^{\otimes k} \right) \bigotimes_{p=1}^m \left( \sum_{j=0}^{d-1} \frac{1}{\sqrt{d}} \ket{j}^{\otimes k} \right) \\
        =& \sum_{j=0}^{d-1} \frac{\alpha_j}{d^{m/2}} \ket{j}^{\otimes k} \sum_{a_{j,1}, \dots, a_{j,m} \in \{0, \dots, d-1\}} \bigotimes_{p=1}^m \ket{a_{j,p}}^{\otimes k}.
    \end{align*}
    Note that there are \( m + 1 \) parts of qudits, each containing \( k \) qudits. The \( 0 \)-th part is in the state \( \sum_{j=0}^{d-1} \alpha_j \ket{j}^{\otimes k} \), while the other parts are in the state \( \sum_{j=0}^{d-1} \frac{1}{\sqrt{d}} \ket{j}^{\otimes k} \). Let \( \ket{a_{j,0}} = \ket{j} \) for \( j = 0, 1, \dots, d-1 \).

    In Step 2, we introduce \( m \) ancillary qudits. For each \( p = 1, 2, \dots, m \), we compute the subtraction modulo \( d \) between the \( (p-1) \)-th part and the \( p \)-th part of qudits, storing the result in the \( p \)-th ancillary qudit. This results in the following state:
    \begin{align*}
        &\sum_{j=0}^{d-1} \frac{\alpha_j}{d^{m/2}} \ket{j}^{\otimes k} \sum_{a_{j,1}, \dots, a_{j,m} \in \{0, \dots, d-1\}} \bigotimes_{p=1}^m \ket{a_{j,p}}^{\otimes k} \\
        &\otimes \ket{a_{j,p-1} -_d a_{j,p}}.
    \end{align*}
    The state of the \( p \)-th ancillary qudit is denoted by \( \ket{a_{j,p-1} -_d a_{j,p}} \). To achieve this, we first apply a \( C \)-\( X_d \) gate controlled by an arbitrary qudit in the \( (p-1) \)-th part to the \( p \)-th ancillary qudit. Then, we apply a \( C \)-\( X^-_d \) gate controlled by an arbitrary qudit in the \( p \)-th part to the \( p \)-th ancillary qudit. This step requires two layers of gates and utilizes \( m \) ancillary qudits.

    In Step 3, we measure all ancillary qudits in the computational basis \( \{\ket{0}, \ket{1}, \dots, \ket{d-1}\} \), obtaining measurement results \( c_1, \dots, c_m \in \{0, 1, \dots, d-1\} \). The quantum state then becomes:
    \begin{align*}
        \sum_{j=0}^{d-1} \alpha_j \ket{\hat{a}_{j,0}}^{\otimes k} \bigotimes_{p=1}^m \ket{\hat{a}_{j,p}}^{\otimes k}.
    \end{align*}
    Note that \( \ket{\hat{a}_{j,0}} = \ket{j} \) for \( j = 0, 1, \dots, d-1 \). For \( p = 1, 2, \dots, m \) and \( j = 0, 1, \dots, d-1 \), the following holds:
    \begin{align*}
        \sum_{q=1}^p c_q = \sum_{q=1}^p \left( \hat{a}_{j,q-1} - \hat{a}_{j,q} \right) = \hat{a}_{j,0} - \hat{a}_{j,q} \mod{d}.
    \end{align*}
    Thus, we have:
    \begin{align*}
        j = \hat{a}_{j,0} = \hat{a}_{j,p} + \sum_{q=1}^p c_q \mod{d}.
    \end{align*}

    In Step 4, for each \( p = 1, 2, \dots, m \), if \( \sum_{q=1}^p c_q = b_p \), we apply an \( X_{+b_p} \) gate to each qudit in the \( p \)-th part, resulting in the desired state:
    \begin{equation*}
        \sum_{j=0}^{d-1} \alpha_j \ket{j}^{\otimes k(m+1)}.
    \end{equation*}

    \begin{figure}
        \centering
        \begin{minipage}{0.9\textwidth}
        \Qcircuit @C=0.85em @R=1em @!R{
        \lstick{\ket{\psi}} & \qw & \ctrl{1} & \qw & \qw & \qw & \qw & \qw \\
        \lstick{\ket{0}} & \qw & \gate{X_d} \qwx[1] & \qw & \qw & \qw & \qw & \qw \\
        \lstick{\ket{0}} & \qw & \gate{X_d} & \ctrl{1} & \qw & \qw & \qw & \qw \\
        \lstick{\ket{0}} & \qw & \qw & \gate{X_d} & \gate{X^-_d} & \meter & \ustick{_{c_1} \ \ \ \ \ \ \ \ } \cw \cwx[1] &  \\
        \lstick{\ket{0}} & \gate{H_d} & \ctrl{1} & \qw & \ctrl{-1} & \qw & \gate{X_{+c_1}} \cwx[1] & \qw \\
        \lstick{\ket{0}} & \qw & \gate{X_d} \qwx[1] & \qw & \qw & \qw & \gate{X_{+c_1}} \cwx[1] & \qw \\
        \lstick{\ket{0}} & \qw & \gate{X_d} & \ctrl{1} & \qw & \qw & \gate{X_{+c_1}} \cwx[1] & \qw \\
        \lstick{\ket{0}} & \qw & \qw & \gate{X_d} & \gate{X^-_d} & \meter & \ustick{_{c_2} \ \ \ \ \ \ \ \ } \cw \cwx[1] &  \\
        \lstick{\ket{0}} & \gate{H_d} & \ctrl{1} & \qw & \ctrl{-1} & \qw & \gate{X_{+(c_1 + c_2)}} \cwx[1] & \qw \\
        \lstick{\ket{0}} & \qw & \gate{X_d} \qwx[1] & \qw & \qw & \qw & \gate{X_{+(c_1 + c_2)}} \cwx[1] & \qw \\
        \lstick{\ket{0}} & \qw & \gate{X_d} & \qw & \qw & \qw & \gate{X_{+(c_1 + c_2)}} & \qw \\
        }
        \end{minipage}
        \caption{Given a qudit in the state \( \ket{\psi} = \sum_{j=0}^{d-1} \alpha_j \ket{j} \), this circuit prepares the state \( \sum_{j=0}^{d-1} \alpha_j \ket{j}^{\otimes 9} \) using two ancillary qudits.}
        \label{fig:fan-out2_d}
    \end{figure}
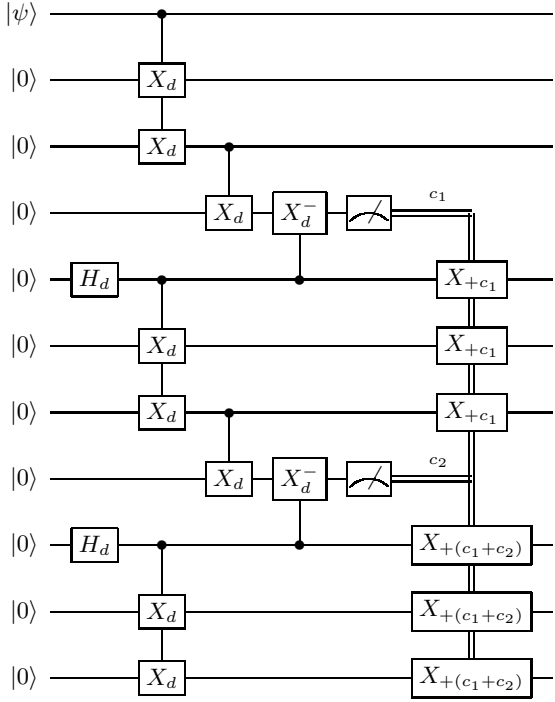

    An example for \( k = 3 \) and \( m = 2 \) is shown in Fig.~\ref{fig:fan-out2_d}. The entire circuit has constant depth. By employing \( m \) additional ancillary qudits, we can prepare an \( n = k(m + 1) \)-qudit state for an arbitrary constant \( k \). To satisfy the requirement, we set \( k \geq \frac{nc}{n + c} \). This circuit requires one layer of intermediate measurements.
\end{proof}

\begin{lemma}
\label{lem:recover_d}
    Given $n$ $d$-dimensional qudits in the state $\sum_{j=0}^{d - 1} \alpha_j\ket{j}^{\otimes n}$, a single qudit in the state $\sum_{j=0}^{d - 1} \alpha_j\ket{j}$ can be obtained using a constant-depth quantum circuit. Simultaneously, all other $n - 1$ qudits can be measured and released. This circuit requires only one layer of intermediate measurements.
\end{lemma}

\begin{proof}
    Initially, we have $n$ qudits in the state:
    \begin{equation*}
        \sum_{j=0}^{d-1} \alpha_j \ket{j}^{\otimes n}.
    \end{equation*}
    
    \textbf{Step 1:} Apply the $H_d$ gate to the last $n - 1$ qudits, resulting in the following state, where $w_d = e^{2i\pi/d}$:
    \begin{equation*}
        \sum_{j=0}^{d-1} \alpha_j \ket{j} \left( \sum_{k=0}^{d-1} \frac{w_d^{jk}}{\sqrt{d}} \ket{k} \right)^{\otimes (n-1)}.
    \end{equation*}
    
    \textbf{Step 2:} Measure the last $n - 1$ qudits, obtaining measurement results $c_1, c_2, \dots, c_{n - 1}$. The remaining state is:
    \begin{equation*}
        \sum_{j=0}^{d-1} \alpha_j w_d^{j\sum_{k=1}^{n-1} c_k} \ket{j}.
    \end{equation*}
    
    \textbf{Step 3:} Classically compute $c = \sum_{k = 1}^{n - 1} c_k \bmod{d}$. Then, apply a $Z^{-c}$ gate to the remaining qudit to obtain the desired state:
    \begin{equation*}
        \sum_{j=0}^{d-1} \alpha_j \ket{j}.
    \end{equation*}
    
    The other $n - 1$ qudits have been measured and can be reset for reuse in subsequent circuits. This circuit utilizes one layer of intermediate measurements in Step 2.
\end{proof}

\begin{theorem}
\label{the:fanout_d}
    The quantum fan-out gate $F_{n,d}$ can be implemented using a constant-depth quantum circuit with $n/c$ ancillary qudits, where $c \in \mathbb{Z}^{+}$ is a constant. This circuit requires one layer of intermediate measurements.
\end{theorem}

\begin{proof}
    Assume $n = mp$, where $m, p \in \mathbb{Z}^{+}$ and $p$ is a constant. We initialize $2m$ ancillary qudits to $\ket{0}$, denoted as $\ket{0}_1, \ket{0}_2, \dots, \ket{0}_{2m}$. The initial state is:
    \begin{equation*}
        \ket{x_0}\ket{0}_1\left(\bigotimes_{j=1}^{m-1} \ket{0}_{2j}\ket{0}_{2j+1} \left(\bigotimes_{k=1}^{p} \ket{x_{j,k}} \right) \right)\ket{0}_{2m} \bigotimes_{k=1}^{p} \ket{x_{m,k}}.
    \end{equation*}
    
    \textbf{Step 1:} Apply $H_d$ gates to the $2j$-th ancillary qudits for $j = 1, 2, \dots, m$, resulting in:
    \begin{align*}
        & \frac{1}{2^{m/2}} \sum_{a_1,\dots,a_m \in \{0,\dots,d-1\}} \ket{x_0}\ket{0}_1\\
        & \otimes \left(\bigotimes_{j=1}^{m-1} \ket{a_j}_{2j} \ket{0}_{2j+1} \left(\bigotimes_{k=1}^{p} \ket{x_{j,k}} \right) \right) \\
        & \otimes \ket{a_{m}}_{2m-1} \bigotimes_{k=1}^{p} \ket{x_{m,k}}.
    \end{align*}
    
    \textbf{Step 2:} Use controlled-$X_d$ gates ($C$-$X_d$), $F_{p,d}$, and $F_{p + 1,d}$ gates to obtain the following state. This step requires a constant-depth quantum circuit:
    \begin{align*}
        & \frac{1}{2^{m/2}} \sum_{a_1,\dots,a_m \in \{0,\dots,d-1 \}} \ket{x_0}\ket{x_0}_1 \\
        & \otimes \left(\bigotimes_{j=1}^{m-1} \ket{a_j}_{2j}\ket{a_j}_{2j+1} \left(\bigotimes_{k=1}^{p} \ket{x_{j,k} +_d a_j} \right) \right) \\
        & \otimes \ket{a_{m}}_{2m} \bigotimes_{k=1}^{p} \ket{x_{m,k} +_d a_{m}}.
    \end{align*}
    
    \textbf{Step 3:} Apply one layer of controlled-$X_{-d}$ gates ($C$-$X_{-d}$) to obtain:
    \begin{align*}
        & \frac{1}{2^{m/2}} \sum_{a_1,\dots,a_m \in \{0,\dots,d-1 \}} \ket{x_0}\ket{x_0 -_d a_1}_1 \\
        & \otimes \left(\bigotimes_{j=1}^{m-1} \ket{a_j}_{2j}\ket{a_j -_d a_{j+1}}_{2j+1} \left(\bigotimes_{k=1}^{p} \ket{x_{j,k} +_d a_j} \right) \right)\\
        & \otimes \ket{a_{m}}_{2m} \bigotimes_{k=1}^{p} \ket{x_{m,k} +_d a_{m}}.
    \end{align*}
    
    \textbf{Step 4:} Measure the $(2j-1)$-th ancillary qudits for $j = 1, 2, \dots, m$, obtaining measurement results $c_{2j-1}$ for $j = 1, 2, \dots, m$. The remaining state is:
    \begin{equation*}
        \ket{x_0} \bigotimes_{j=1}^{m} \left( \ket{\hat{a}_j}_{2j} \bigotimes_{k=1}^{p} \ket{x_{j,k} +_d \hat{a}_j} \right).
    \end{equation*}
    The following relationships hold (all computations are performed modulo \( d \)):
    \begin{align*}
        & c_1 = x_0 -_d \hat{a}_1, \\
        & c_{2j-1} = \hat{a}_{j-1} -_d \hat{a}_{j}, \quad j = 2, 3, \dots, m, \\
        & b_j = \sum_{k=1}^{j} c_{2k-1} = x_0 -_d \hat{a}_{j}, \quad j = 1, 2, \dots, m.
    \end{align*}
    For $j = 1, 2, \dots, m$, compute $b_j$ classically and apply $X_{+b_j}$ gates to the qudits in the state $\bigotimes_{k=1}^{p} \ket{x_{j,k} +_d \hat{a}_j}$, resulting in:
    \begin{equation*}
        \ket{x_0} \bigotimes_{j=1}^{m} \left( \ket{\hat{a}_j}_{2j} \bigotimes_{k=1}^{p} \ket{x_{j,k} +_d x_0} \right).
    \end{equation*}
    
    \textbf{Step 5:} Apply $H_d$ gates to the $2j$-th ancillary qudits for $j = 1, 2, \dots, m$, then measure these qudits to obtain results $c_{2j}$ for $j = 1, 2, \dots, m-1$. The resulting state is:
    \begin{equation*}
        w_d^{\sum_{j=1}^{m}\hat{a}_{j}c_{2j}}\ket{x_0} \bigotimes_{j=1}^{m} \bigotimes_{k=1}^{p} \ket{x_{j,k} +_d x_0}, \quad w_d = e^{2i\pi/d}.
    \end{equation*}
    Classically compute $t = \sum_{j=1}^{m} c_{2j}$ and apply a $Z^{-t}$ gate to $\ket{x_0}$, yielding:
    \begin{equation*}
        w_d^{\sum_{j=1}^{m-1}(\hat{a}_{j} -_d x_0)c_{2j}}\ket{x_0} \bigotimes_{j=1}^{m} \bigotimes_{k=1}^{p} \ket{x_{j,k} +_d x_0}.
    \end{equation*}
    Since $x_0 -_d \hat{a}_{j} = \sum_{k=1}^{j} c_{2k-1}$ is a constant, the term $w_d^{\sum_{j=1}^{m-1}(\hat{a}_{j} -_d x_0)c_{2j}}$ is a global phase and can be omitted. Thus, the final state is:
    \begin{equation*}
        \ket{x_0} \bigotimes_{j=1}^{m} \bigotimes_{k=1}^{p} \ket{x_{j,k} +_d x_0}.
    \end{equation*}
    The entire circuit has constant depth, and one layer of intermediate measurements is sufficient for Steps 4 and 5. If fewer than $n/c$ ancillary qudits are desired, we can set $m \leq \frac{n}{2c}$.
\end{proof}

\section{Constant-Depth Circuit for Quantum State Preparation}
\label{app:qsp}

\subsection{Constant-Depth Circuit for the $n$-Toffoli Gate}

The $n$-Toffoli gate (also known as the $C^n$-$X$ gate) performs the following transformation, where $x_1, \dots, x_n, y \in \{0, 1\}$:
\begin{equation*}
    \ket{x_1, \dots, x_n}\ket{y} \to \ket{x_1, \dots, x_n}\ket{y \oplus \prod_{j=1}^{n}x_j}.
\end{equation*}

In \cite{takahashi2016collapse}, a constant-depth exact implementation of the $n$-Toffoli gate is proposed, utilizing quantum fan-out gates. Building on this result, a constant-depth dynamic circuit implementation of the $n$-Toffoli gate is directly obtained as described in \cite{buhrman2024state}.

\begin{theorem}
\label{the:toffoli}
    The \( n \)-Toffoli gate can be implemented using a constant-depth quantum circuit with a circuit size of \( O(n \log n) \) and \( O(n \log n) \) ancilla. This circuit comprises \( 6 \) layers of intermediate measurements \cite{takahashi2016collapse}.
\end{theorem}

According to \cite{takahashi2016collapse}, the number of ancilla in Theorem~\ref{the:toffoli}, which is $O(n\log n)$, can be reduced to $O(n\log^{(c)} n)$ for any integer constant $c > 1$. This reduction comes at the cost of increasing the circuit depth by a constant factor. Here, $\log^{(c)} n$ denotes the $c$-times iterated logarithm. In practice, by setting $c = 5$, the $n$-Toffoli gate can be implemented using a constant-depth circuit with $O(n)$ ancilla. This is because, for all practical values of $n$, $\log^{(5)} n < 10$.

\subsection{Constant-Depth Circuit for Quantum State Preparation}

Our goal is to construct a quantum circuit that prepares the following $n$-qubit state:
\begin{equation*}
    \sum_{j = 0}^{2^n-1} \alpha_{j}e^{i\theta_{j}} \ket{j}, \quad \alpha_{j} \in [0, 1], \quad \theta_{j} \in [0, 2\pi), \quad \sum_{j=0}^{2^n-1} \alpha_j^2 = 1.
\end{equation*}

The construction proceeds in three steps:
\begin{enumerate}
    \item Prepare the state $\sum_{j=0}^{2^n-1} \alpha_{j} \ket{e_{j}}$, where $e_j$ is the one-hot encoding of $j$.
    \item Apply the phase $e^{i\theta_j}$ to each state $\ket{e_j}$.
    \item Transform each $\ket{e_j}$ into $\ket{j}$ to obtain the desired state.
\end{enumerate}

The most technical part of our construction is Lemma~\ref{lem:onehot}.

\begin{lemma}
\label{lem:onehot}
    The following \( 2^n \)-qubit state can be prepared from the initial state \( \ket{0}^{\otimes 2^n} \) using a constant-depth quantum circuit with a size of \( O(n4^n) \). This circuit utilizes \( 22 \) layers of intermediate measurements and \( O(n4^n) \) ancilla.
    \begin{align*}
        &\sum_{j=0}^{2^n-1} \alpha_j \ket{e_{j}}, \quad \text{where} \\
        &\ket{e_0} = \ket{1} \otimes \ket{0}^{2^n-1}, \\
        &\ket{e_j} = \ket{0}^{\otimes j} \ket{1} \ket{0}^{\otimes (2^n-j-1)}, \quad j = 1, 2, \dots, 2^n-1.
    \end{align*}
\end{lemma}

\begin{proof}
    Initially, we have $2^n$ qubits in the state $\ket{0}^{\otimes 2^n}$. For convenience, let $N = 2^n$. We define angles $\theta_1, \theta_2, \dots, \theta_{N-1}$ such that:
    \begin{align*}
        &\sin \theta_1 = \alpha_0, \\
        &\sin \theta_k = \frac{\alpha_{k-1}}{\sqrt{1 - \sum_{j=0}^{k-2} \alpha_j^2}}, \quad k = 2, 3, \dots, N-1.
    \end{align*}
    Through direct calculation, we verify the following relationships:
    \begin{align*}
        &\alpha_0 = \sin \theta_1, \quad \alpha_{N-1} = \prod_{j=1}^{N-1} \cos \theta_j, \\
        &\alpha_{k-1} = \left(\prod_{j=1}^{k-1} \cos \theta_j \right) \sin \theta_k, \quad k = 2, 3, \dots, N-1.
    \end{align*}

    \textbf{Step 1:} Apply the $Ry(2\theta_k)$ gate to the $k$-th qubit for $k = 1, 2, \dots, N-1$. This transforms the state to:
    \begin{align*}
        \left(\bigotimes_{j=1}^{N-1} \left(\cos \theta_j \ket{0} + \sin \theta_j \ket{1} \right) \right) \otimes \ket{0}.
    \end{align*}
    This step requires only one layer of single-qubit gates. Temporarily omitting the $N$-th qubit, we expand the state as:
    \begin{align*}
        &\sum_{\boldsymbol{z} \in \{0,1\}^{N-1}} a_{\boldsymbol{z}} \ket{\boldsymbol{z}}, \quad \boldsymbol{z} = z_1z_2\cdots z_{N-1}, \\
        &a_{\boldsymbol{z}} = \prod_{j=1}^{N-1} \left( (1 - z_j) \cos \theta_j + z_j \sin \theta_j \right).
    \end{align*}
    For simplicity, we denote the state of the $j$-th qubit as $\ket{z_j}$ for $j = 1, 2, \dots, N-1$.

    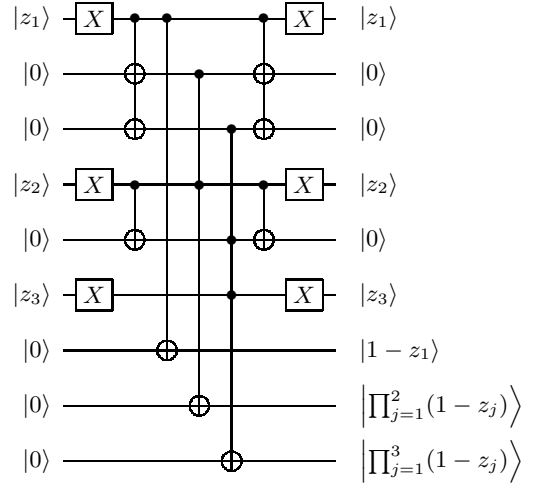
\begin{figure}
        \centering
        \begin{minipage}{0.9\textwidth}
        \Qcircuit @C=0.5em @R=1em @!R{
        \lstick{\ket{z_1}} & \gate{X} & \ctrl{1} & \ctrl{6} & \qw & \qw & \ctrl{1} & \gate{X} & \qw & \rstick{\ket{z_1}} \\
        \lstick{\ket{0}} & \qw & \targ \qwx[1] & \qw & \ctrl{2} & \qw & \targ \qwx[1] & \qw & \qw & \rstick{\ket{0}} \\
        \lstick{\ket{0}} & \qw & \targ & \qw & \qw & \ctrl{2} & \targ & \qw & \qw & \rstick{\ket{0}} \\
        \lstick{\ket{z_2}} & \gate{X} & \ctrl{1} & \qw & \ctrl{4} & \qw & \ctrl{1} & \gate{X} & \qw & \rstick{\ket{z_2}} \\
        \lstick{\ket{0}} & \qw & \targ & \qw & \qw & \ctrl{1} & \targ & \qw & \qw & \rstick{\ket{0}} \\
        \lstick{\ket{z_3}} & \gate{X} & \qw & \qw & \qw & \ctrl{3} & \qw & \gate{X} & \qw & \rstick{\ket{z_3}} \\
        \lstick{\ket{0}} & \qw & \qw & \targ & \qw & \qw & \qw & \qw & \qw & \rstick{\ket{1-z_1}} \\
        \lstick{\ket{0}} & \qw & \qw & \qw & \targ & \qw & \qw & \qw & \qw & \rstick{\ket{\prod_{j=1}^2(1-z_j)}} \\
        \lstick{\ket{0}} & \qw & \qw & \qw & \qw & \targ & \qw & \qw & \qw & \rstick{\ket{\prod_{j=1}^3(1-z_j)}}
        }
        \end{minipage}
        \caption{A $3$-qubit instance of the circuit for Step 2 in Lemma~\ref{lem:onehot}.}
        \label{fig:prefix}
    \end{figure}

    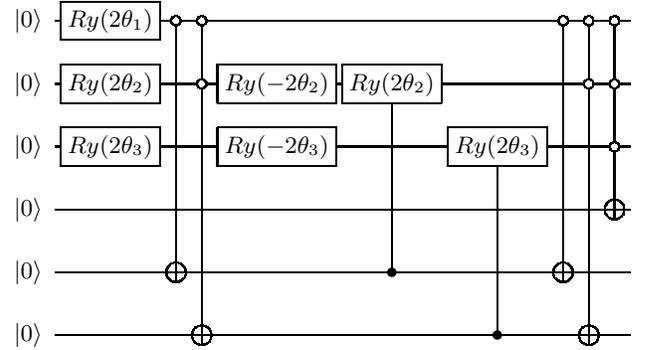
\begin{figure}
        \centering
        \begin{minipage}{0.9\textwidth}
        \Qcircuit @C=0.23em @R=1em @!R{
        \lstick{\ket{0}} & \gate{Ry(2\theta_1)} & \ctrlo{4} & \ctrlo{1} & \qw & \qw & \qw & \ctrlo{4} & \ctrlo{1} & \ctrlo{1} & \qw \\
        \lstick{\ket{0}} & \gate{Ry(2\theta_2)} & \qw & \ctrlo{4} & \gate{Ry(-2\theta_2)} & \gate{Ry(2\theta_2)} & \qw & \qw & \ctrlo{4} & \ctrlo{1} & \qw \\
        \lstick{\ket{0}} & \gate{Ry(2\theta_3)} & \qw & \qw & \gate{Ry(-2\theta_3)} & \qw & \gate{Ry(2\theta_3)} & \qw & \qw & \ctrlo{1} & \qw \\
        \lstick{\ket{0}} & \qw & \qw & \qw & \qw & \qw & \qw & \qw & \qw & \targ & \qw \\
        \lstick{\ket{0}} & \qw & \targ & \qw & \qw & \ctrl{-3} & \qw & \targ & \qw & \qw & \qw \\
        \lstick{\ket{0}} & \qw & \qw & \targ & \qw & \qw & \ctrl{-3} & \qw & \targ & \qw & \qw 
        }
        \end{minipage}
        \caption{A circuit instance for preparing the state $\sum_{j=0}^3 \alpha_j \ket{e_j}$.}
        \label{fig:onehot}
    \end{figure}

    \textbf{Step 2:} Initialize $N - 2$ ancilla to $\ket{0}$. Compute $\prod_{j=1}^k (1 - z_j)$ and store the result in the $k$-th ancilla for $k = 1, 2, \dots, N-2$. The resulting state is:
    \begin{equation*}
        \sum_{\boldsymbol{z} \in \{0,1\}^{N-1}} a_{\boldsymbol{z}} \ket{\boldsymbol{z}} \bigotimes_{k=1}^{N-2} \ket{\prod_{j=1}^k (1 - z_j)}.
    \end{equation*}

    To implement Step 2 in constant depth, we first apply $X$ gates to each qubit in $\ket{\boldsymbol{z}}$ to obtain $\bigotimes_{j=1}^{N-1} \ket{1 - z_j}$. We then copy each $(1 - z_j)$ sufficiently many times, resulting in:
    \begin{equation*}
        \left(\bigotimes_{j=1}^{N-2} \ket{1 - z_j}^{\otimes (N-1-j)} \right) \ket{1 - z_{N-1}} \ket{0}^{\otimes (N-2)}.
    \end{equation*}
    This step can be achieved using a constant-depth circuit with $O(N^2) = O(4^n)$ ancilla, as per Lemma~\ref{lem:GHZ}. Next, we use a constant-depth circuit with $O(N^2 \log N) = O(n4^n)$ ancilla (based on Theorem~\ref{the:toffoli}) to obtain:
    \begin{equation*}
        \left(\bigotimes_{j=1}^{N-2} \ket{1 - z_j}^{\otimes (N-1-j)} \right) \ket{1 - z_{N-1}} \bigotimes_{k=1}^{N-2} \ket{\prod_{j=1}^k (1 - z_j)}.
    \end{equation*}
    Using Lemma~\ref{lem:recover}, we delete the redundant copies of $(1 - z_j)$. Finally, we apply a layer of $X$ gates to recover $\ket{z_j}$, yielding the desired state:
    \begin{equation*}
        \ket{\boldsymbol{z}} \bigotimes_{k=1}^{N-2} \ket{\prod_{j=1}^k (1 - z_j)}.
    \end{equation*}
    An example circuit for $N - 2 = 3$ is shown in Fig.~\ref{fig:prefix}. Quantum fan-out gates are used to represent the application of Lemma~\ref{lem:GHZ} and Lemma~\ref{lem:recover}, simplifying the explanation.

    \textbf{Step 3:} Apply the $Ry(-2\theta_k)$ gate to the $k$-th qubit for $k = 2, 3, \dots, N-1$. This step is similar to Step 1, but with the parameter of the $Ry$ gate negated. The resulting state is:
    \begin{align*}
        &\sin \theta_1 \ket{1} \ket{0}^{\otimes (N-2)} \otimes \ket{0}^{\otimes (N-2)} \\
        &+ \prod_{j=1}^{N-1} \cos \theta_j \ket{0} \bigotimes_{j=2}^{N-1} (\cos \theta_j \ket{0} - \sin \theta_j \ket{1}) \ket{1}^{\otimes (N-2)} \\
        &+ \sum_{k=2}^{N-1} \prod_{j=1}^{k-1} \cos \theta_j \ket{0} \bigotimes_{j=2}^{k-1} (\cos \theta_j \ket{0} - \sin \theta_j \ket{1}) \\
        & \otimes \sin \theta_{k} (\cos \theta_k \ket{0} + \sin \theta_k \ket{1}) \ket{0}^{\otimes (N-1-k)} \\
        & \otimes \ket{1}^{\otimes (k-1)} \ket{0}^{\otimes (N-k-1)}.
    \end{align*}

    \textbf{Step 4:} Apply controlled-$Ry(2\theta_k)$ gates for $k = 1, 2, \dots, N-2$. The control qubit is the $k$-th ancilla, which stores $\ket{\prod_{j=1}^k (1 - z_j)}$. When the control qubit is $\ket{1}$, the $Ry(2\theta_k)$ gate is applied to the $(k + 1)$-th qubit. This step can be implemented using a constant-depth circuit without additional ancilla, as described in \cite{nielsen2010quantum}. The resulting state is:
    \begin{align*}
        &\sin \theta_1 \ket{1} \ket{0}^{\otimes (N-2)} \otimes \ket{0}^{\otimes (N-2)} \\
        &+ \prod_{j=1}^{N-1} \cos \theta_j \ket{0}^{\otimes (N-1)} \ket{1}^{\otimes (N-2)} \\
        &+ \sum_{k=2}^{N-1} \prod_{j=1}^{k-1} \cos \theta_j \ket{0}^{\otimes (k-1)} \sin \theta_k \ket{1} \ket{0}^{\otimes (N-1-k)} \\
        &\quad \otimes \ket{1}^{\otimes (k-1)} \ket{0}^{\otimes (N-k-1)} \\
        =& \alpha_0 \ket{1} \ket{0}^{\otimes (N-2)} \otimes \ket{0}^{\otimes (N-2)} \\
        &+ \alpha_{N-1} \ket{0}^{\otimes (N-1)} \otimes \ket{1}^{\otimes (N-2)} \\
        &+ \sum_{k=2}^{N-1} \alpha_{k-1} \ket{0}^{\otimes (k-1)} \ket{1} \ket{0}^{\otimes (N-1-k)} \\
        &\quad \otimes \ket{1}^{\otimes (k-1)} \ket{0}^{\otimes (N-k-1)}.
    \end{align*}

    \textbf{Step 5:} Recover the $N - 2$ ancilla by repeating Step 2. After this step, we ignore the recovered ancilla and add the $N$-th qubit $\ket{0}$. The resulting state is:
    \begin{align*}
        &\alpha_0 \ket{1} \ket{0}^{\otimes (N-1)} \\
        &+ \sum_{k=2}^{N-1} \alpha_{k-1} \ket{0}^{\otimes (k-1)} \ket{1} \ket{0}^{\otimes (N-k)} + \alpha_{N-1} \ket{0}^{\otimes N}.
    \end{align*}

    \textbf{Step 6:} Apply an $(N - 1)$-qubit controlled $X$ gate. When the first $N - 1$ qubits are in the state $\ket{0}^{\otimes (N - 1)}$, an $X$ gate is applied to the $N$-th qubit. This step can be implemented using a constant-depth circuit with $O(N \log N) = O(n2^n)$ ancilla, as per Theorem~\ref{the:toffoli}. The final state is:
    \begin{align*}
        &\alpha_0 \ket{1} \ket{0}^{\otimes (N-1)} + \sum_{k=2}^{N-1} \alpha_{k-1} \ket{0}^{\otimes (k-1)} \ket{1} \ket{0}^{\otimes (N-k)} \\
        &+ \alpha_{N-1} \ket{0}^{\otimes (N-1)} \ket{1} \\
        =& \sum_{j=0}^{N-1} \alpha_j \ket{e_j}.
    \end{align*}

    An example circuit for preparing the state $\sum_{j=0}^3 \alpha_j \ket{e_j}$ is shown in Fig.~\ref{fig:onehot}. The hollow circles in the figure indicate that the target qubit undergoes a quantum gate when the control qubit is in the state $\ket{0}$. Gates similar to the CNOT and Toffoli gates are used to implement Steps 2 and 5, as illustrated in Fig.~\ref{fig:prefix}.

    The entire quantum circuit has constant depth and uses $O(n4^n)$ ancilla. Steps 2 and 5 each require $8$ layers of intermediate measurements, while Step 6 requires $6$ layers. Thus, the total number of intermediate measurement layers is $22$. Since the circuit depth is constant, the circuit size is bounded by \( O(n + m) \) when the circuit utilizes \( m \) ancilla. This concludes the proof.
\end{proof}

\begin{lemma}
\label{lem:tobinary}
    There exists a constant-depth quantum circuit $C$ with $O(n2^n\log n)$ ancilla and $9$ layers of intermediate measurements, such that for any $N$-qubit state $\sum_{j=0}^{N-1} \alpha_j \ket{e_j}$, where $N = 2^n$, $\alpha_j \in \mathbb{C}$, $\sum_{j=0}^{N-1} |\alpha_j|^2 = 1$, and
    \begin{align*}
        \ket{e_0} &= \ket{1}\ket{0}^{\otimes (N-1)}, \\
        \ket{e_j} &= \ket{0}^{\otimes j}\ket{1}\ket{0}^{\otimes (N-j-1)}, \quad j = 1, 2, \dots, N-1,
    \end{align*}
    the circuit $C$ satisfies:
    \begin{equation*}
        C\sum_{j=0}^{N-1} \alpha_j \ket{e_j} = \sum_{j=0}^{N-1} \alpha_j \ket{j}\ket{0}^{\otimes (N-n)}.
    \end{equation*}
\end{lemma}

\begin{proof}
    We provide a concise proof here, as a similar result is presented and utilized in \cite{allcock2023constant}, albeit not formalized as a theorem. Our proof combines Lemma 28 from \cite{sun2023asymptotically} and Theorem~\ref{the:fanout}.

    Initially, we have the $N$-qubit state $\sum_{j=0}^{N-1} \alpha_j \ket{e_j}$.

    \textbf{Step 1:} We aim to transform the state $\sum_{j=0}^{N-1} \alpha_j \ket{e_j}\ket{0}^{\otimes n}$ into:
    \begin{equation*}
        \sum_{j=0}^{N-1} \alpha_j \ket{e_j}\ket{j}.
    \end{equation*}
    To achieve this, we construct a circuit as follows. For each $j = 0, 1, \dots, N-1$, we apply at most $n$ CNOT gates. The control qubit is the $(j + 1)$-th input qubit (corresponding to the input state $\ket{e_j}$), and the target qubits correspond to the positions of $1$ in the binary representation of $j$. These CNOT gates encode $\ket{j}$ into the $n$ ancilla when the input state is $\ket{e_j}$. To implement this in constant depth, we first copy each input qubit at most $n$ times using a constant-depth circuit with $O(n2^n)$ ancilla, as per Lemma~\ref{lem:GHZ}. Since each ancilla is involved in $2^{n-1}$ CNOT gates as the target, and the control qubits are distinct, these gates can be executed in parallel using a constant-depth circuit with $O(N)$ additional ancilla, according to Corollary~\ref{cor:parity}. The total number of ancilla used is $O(n2^n)$.

    \textbf{Step 2:} We transform the state into:
    \begin{equation*}
        \sum_{j=0}^{N-1} \alpha_j \ket{0}^{\otimes N}\ket{j}.
    \end{equation*}
    To achieve this, we construct a circuit that maps $\sum_{j=0}^{N-1} \alpha_j \ket{0}^{\otimes N}\ket{j}$ to $\sum_{j=0}^{N-1} \alpha_j \ket{e_j}\ket{j}$. This circuit consists of $N = 2^n$ multi-controlled $X$ gates. For each $j = 0, 1, \dots, N-1$, when the $n$ ancilla are in state $\ket{j}$, we apply an $X$ gate to the $(j + 1)$-th input qubit, transforming $\ket{0}^{\otimes N}$ into $\ket{e_j}$. To implement this in constant depth, we first copy each ancilla $N = 2^n$ times using a constant-depth circuit with $O(n2^n)$ ancilla, as per Lemma~\ref{lem:GHZ}. The multi-controlled $X$ gates can then be executed in parallel using a constant-depth circuit with $O(n2^n\log n)$ ancilla, according to Theorem~\ref{the:toffoli}.

    \textbf{Step 3:} We swap the $n$ ancilla storing $\ket{j}$ to the first $n$ qubits, resulting in the desired state:
    \begin{equation*}
        \sum_{j=0}^{N-1} \alpha_j \ket{j}.
    \end{equation*}
    This step requires one layer of swap gates and no additional ancilla.

    Step 1 requires $2$ layers of intermediate measurements, and Step 2 requires $7$ layers. The entire circuit thus has $9$ layers of intermediate measurements. This completes the proof.
\end{proof}

\begin{theorem}
\label{the:statepre}
    An arbitrary \( n \)-qubit quantum state of the form \( \sum_{j=0}^{2^n-1} \alpha_j e^{i\theta_j} \ket{j} \) can be prepared from the initial state \( \ket{0}^{\otimes n} \) using a constant-depth quantum circuit. The circuit has a size of \( O(n4^n) \), utilizes \( O(n4^n) \) ancilla, and requires \( 31 \) layers of intermediate measurements.
\end{theorem}

\begin{proof}
    In Step 1, we prepare the state \( \sum_{j=0}^{2^n - 1} \alpha_{j} \ket{e_j} \) using Lemma~\ref{lem:onehot}.

In Step 2, for \( j = 0, 1, \dots, 2^n - 1 \), we apply a \( Z(\theta_j) \) gate to the \( (j+1) \)-th qubit, resulting in the state:
\[
\sum_{j=0}^{2^n - 1} \alpha_{j} e^{i\theta_j} \ket{e_j}.
\]

In Step 3, we transform the state into \( \sum_{j=0}^{2^n - 1} \alpha_{j} e^{i\theta_j} \ket{j} \) using Lemma~\ref{lem:tobinary}.

The entire circuit has a constant depth. The circuit size and ancilla count are bounded by \( O(n4^n) \), and the circuit requires \( 31 \) layers of intermediate measurements, as specified in Lemma~\ref{lem:onehot} and Lemma~\ref{lem:tobinary}.
\end{proof}

\subsection{Constant-Depth Circuit for Sparse Quantum State Preparation}

For an $n$-qubit sparse quantum state $\ket{\psi}$, only a polynomial number of basis states have non-zero coefficients. Such sparse states can be prepared with lower resource costs, as described in the following theorem.

\begin{theorem}
\label{the:sparse}
    For any $n$-qubit sparse quantum state $\ket{\psi}$ with $s$ non-zero coefficients, there exists a constant-depth quantum circuit $C$ that uses $O(s^2 \log n)$ ancilla and $31$ layers of intermediate measurements, such that:
    \begin{equation*}
        C \ket{0}^{\otimes n} = \ket{\psi}.
    \end{equation*}
\end{theorem}

\begin{proof}
    Let $\ket{\psi} = \sum_{j \in S} \alpha_j \ket{j}$, where $S \subset \{0,1\}^n$ is a set of size $s = |S| = O(n^c)$ for some constant $c$. For all $j \in S$, $\alpha_j \in \mathbb{C}$ and $\sum_{j \in S} |\alpha_j|^2 = 1$. We define the following basis states:
    \begin{align*}
        \ket{e_0} &= \ket{1}\ket{0}^{\otimes (s-1)}, \\
        \ket{e_j} &= \ket{0}^{\otimes j}\ket{1}\ket{0}^{\otimes (s-j-1)}, \quad j = 1, 2, \dots, s-1.
    \end{align*}
    Let $\{\ket{\hat{e}_j} \mid j \in S\}$ be a relabeling of the set $\{\ket{e_j} \mid j = 0, 1, \dots, s-1\}$. Using Lemma~\ref{lem:onehot}, we can prepare the state:
    \begin{equation*}
        \sum_{j \in S} |\alpha_j| \ket{\hat{e}_j}.
    \end{equation*}
    According to Lemma~\ref{lem:onehot}, this step requires a constant-depth circuit with $O(s^2 \log s) = O(s^2 \log n)$ ancilla.

    Next, for each $j \in S$, we apply a $Z(\arg(\alpha_j))$ gate to the qubit in state $\ket{1}$ within $\ket{\hat{e}_j}$. This transforms $|\alpha_j| \ket{\hat{e}_j}$ into $\alpha_j \ket{\hat{e}_j}$, resulting in the state:
    \begin{equation*}
        \sum_{j \in S} \alpha_j \ket{\hat{e}_j}.
    \end{equation*}

    Finally, we map each $\ket{\hat{e}_j}$ to $\ket{j}$ using Lemma~\ref{lem:tobinary}. This step requires a constant-depth circuit with $O(ns)$ ancilla and yields the desired state:
    \begin{equation*}
        \ket{\psi} = \sum_{j \in S} \alpha_j \ket{j}.
    \end{equation*}
    The entire circuit has $31$ layers of intermediate measurements, similar to Theorem~\ref{the:statepre}.
\end{proof}

\section{Constant-Depth Circuit for Controlled Quantum State Preparation}
\label{app:cqsp}

\begin{lemma}
\label{lem:toonehot}
    There exists a constant-depth quantum circuit $C$ with $O(n2^n \log n)$ ancilla and $9$ layers of intermediate measurements. For any $n$-qubit state $\sum_{j=0}^{2^n-1} \alpha_j \ket{j}$, where $\alpha_j \in \mathbb{C}$ and $\sum_{j=0}^{2^n-1} |\alpha_j|^2 = 1$, the circuit $C$ satisfies:
    \begin{equation*}
        C \left( \sum_{j=0}^{2^n-1} \alpha_j \ket{j} \ket{0}^{\otimes (2^n - n)} \right) = \sum_{j=0}^{2^n-1} \alpha_j \ket{e_j},
    \end{equation*}
    where
    \begin{align*}
        \ket{e_0} &= \ket{1} \ket{0}^{\otimes (2^n - 1)}, \\
        \ket{e_j} &= \ket{0}^{\otimes j} \ket{1} \ket{0}^{\otimes (2^n - j - 1)}, \quad j = 1, 2, \dots, 2^n - 1.
    \end{align*}
\end{lemma}

\begin{proof}
    The proof of Lemma~\ref{lem:toonehot} follows directly from the proof of Lemma~\ref{lem:tobinary}. Initially, we have $2^n$ qubits in the state $\sum_{j=0}^{2^n-1} \alpha_j \ket{j} \ket{0}^{\otimes (2^n - n)}$. 

    \textbf{Step 1:} Using Step 2 of Lemma~\ref{lem:tobinary}, we transform the state into:
    \begin{equation*}
        \sum_{j=0}^{2^n-1} \alpha_j \ket{j} \ket{e_j}.
    \end{equation*}

    \textbf{Step 2:} Using Step 1 of Lemma~\ref{lem:tobinary}, we further transform the state into:
    \begin{equation*}
        \sum_{j=0}^{2^n-1} \alpha_j \ket{0}^{\otimes n} \ket{e_j}.
    \end{equation*}

    \textbf{Step 3:} Finally, we swap qubits as in Step 3 of Lemma~\ref{lem:tobinary} to obtain the desired state:
    \begin{equation*}
        \sum_{j=0}^{2^n-1} \alpha_j \ket{e_j}.
    \end{equation*}
\end{proof}

We now prove in Lemma~\ref{lem:CQSP} that controlled quantum state preparation can be achieved without an asymptotic increase in the number of ancilla.

\begin{lemma}
\label{lem:CQSP}
    For an arbitrary \( n \)-qubit state \( \sum_{j = 0}^{2^n - 1} \alpha_j \ket{j} \) and a one-qubit state \( a\ket{0} + b\ket{1} \), where \( a, b, \alpha_0, \dots, \alpha_{2^n - 1} \in \mathbb{C} \), a constant-depth quantum circuit \( C \) of size \( O(n4^n) \) can perform controlled state preparation. This circuit employs \( O(n4^n) \) ancilla and requires \( 33 \) layers of intermediate measurements.
    \begin{equation*}
        C \left( (a\ket{0} + b\ket{1}) \ket{0}^{\otimes n} \right) = a\ket{0}^{\otimes (n+1)} + b\ket{1} \sum_{j=0}^{2^n-1} \alpha_j \ket{j}.
    \end{equation*}
\end{lemma}

\begin{proof}
    The proof builds on Theorem~\ref{the:statepre}, which prepares an arbitrary quantum state using Lemma~\ref{lem:onehot} and Lemma~\ref{lem:tobinary}. We extend this construction to a controlled version.

    Initially, we have $n + 1$ qubits in the state $(a\ket{0} + b\ket{1}) \ket{0}^{\otimes n}$.

    \textbf{Controlled Version of Lemma~\ref{lem:onehot}:} The circuit in Lemma~\ref{lem:onehot} consists of six steps. We modify these steps to incorporate a control qubit:
    \begin{itemize}
        \item For Steps 2 and 5, no changes are needed as their effects cancel out.
        \item For Steps 1, 3, and 4, we first copy the control qubit $2^n$ times using Lemma~\ref{lem:GHZ}. These steps are then modified to their controlled versions, which still require constant depth.
        \item Step 6 involves an $(2^n - 1)$-controlled $X$ gate and single-qubit gates. We add a control qubit to create a $2^n$-controlled $X$ gate, maintaining constant depth.
    \end{itemize}
    The controlled version of Lemma~\ref{lem:onehot} requires $23$ layers of intermediate measurements (including an additional layer for copying the control qubit) and yields the state:
    \begin{equation*}
        a \ket{0}^{\otimes 2^n} \ket{0}^{\otimes 2^n} + b \ket{1}^{\otimes 2^n} \sum_{j=0}^{2^n-1} |\alpha_j| \ket{e_{j}}.
    \end{equation*}

    \textbf{Phase Application:} We apply a layer of controlled-$Z(\theta)$ gates to transform $|\alpha_j|$ into $\alpha_j$. This step requires constant depth due to the availability of sufficient copies of the control qubit. Using Lemma~\ref{lem:recover}, we remove redundant copies, resulting in:
    \begin{equation*}
        a \ket{0} \ket{0}^{\otimes 2^n} + b \ket{1} \sum_{j=0}^{2^n-1} \alpha_j \ket{e_{j}}.
    \end{equation*}

    \textbf{Controlled Version of Lemma~\ref{lem:tobinary}:} To transform $\ket{0}^{\otimes 2^n}$ into $\ket{e_0}$ when the control qubit is $\ket{0}$, we apply one CNOT gate and two $X$ gates, yielding:
    \begin{equation*}
        a \ket{0} \ket{e_0} + b \ket{1} \sum_{j=0}^{2^n-1} \alpha_j \ket{e_j}.
    \end{equation*}
    Finally, we apply Lemma~\ref{lem:tobinary} to obtain the desired state:
    \begin{equation*}
        a \ket{0}^{\otimes (n+1)} + b \ket{1} \sum_{j=0}^{2^n-1} \alpha_j \ket{j}.
    \end{equation*}

    The entire circuit has constant depth, uses $O(n4^n)$ ancilla, and requires $33$ layers of intermediate measurements. Since the circuit depth is constant, the circuit size is bounded by \( O(n + m) \) when the circuit utilizes \( m \) ancilla.
\end{proof}

\begin{lemma}
\label{lem:nRy}
    Controlled-$Ry(\theta_j)$ gates for arbitrary angles $\theta_1, \dots, \theta_n$ (where $\theta_j \in [0, 2\pi)$) sharing a common target qubit can be implemented using a constant-depth quantum circuit with $O(n)$ ancilla. This circuit requires two layers of intermediate measurements. The transformation is as follows:
    \begin{align*}
        &\ket{t} \otimes \ket{x_1, \dots, x_n} \to Ry\left(\sum_{j=1}^n x_j \theta_j\right) \ket{t} \otimes \ket{x_1, \dots, x_n}, \\
        &x_1, \dots, x_n \in \{0, 1\}.
    \end{align*}
\end{lemma}
\begin{proof}
    Since $Ry(\theta)$ gates commute with each other, this lemma is a direct consequence of Theorem 3.2 in \cite{hoyer2005quantum} and Theorem~\ref{the:fanout}.
\end{proof}

\begin{lemma}
\label{lem:nZ}
    Controlled-$Z(\theta_j)$ gates for arbitrary angles $\theta_1, \dots, \theta_n$ (where $\theta_j \in [0, 2\pi)$) sharing a common target qubit can be implemented using a constant-depth quantum circuit with $O(n)$ ancilla. This circuit requires two layers of intermediate measurements. The transformation is as follows:
    \begin{align*}
        &\ket{t} \otimes \ket{x_1, \dots, x_n} \to Z\left(\sum_{j=1}^n x_j \theta_j\right) \ket{t} \otimes \ket{x_1, \dots, x_n}, \\
        &x_1, \dots, x_n \in \{0, 1\}.
    \end{align*}
\end{lemma}
\begin{proof}
    Since $Z(\theta)$ gates commute with each other, this lemma is a direct consequence of Theorem 3.2 in \cite{hoyer2005quantum} and Theorem~\ref{the:fanout}.
\end{proof}

\begin{lemma}
\label{lem:CCnRy}
    A constant-depth quantum circuit with $O(n)$ ancilla can implement the following transformation. This circuit requires four layers of intermediate measurements:
    \begin{align*}
        &\ket{t} \ket{x_1, \dots, x_n} \ket{c} \to R_y\left(c \sum_{j=1}^n x_j \theta_j\right) \ket{t} \ket{x_1, \dots, x_n} \ket{c}, \\
        &\theta_1, \dots, \theta_n \in [0, 2\pi); \quad x_1, \dots, x_n \in \{0, 1\}; \quad c \in \{0, 1\}.
    \end{align*}
\end{lemma}
\begin{proof}
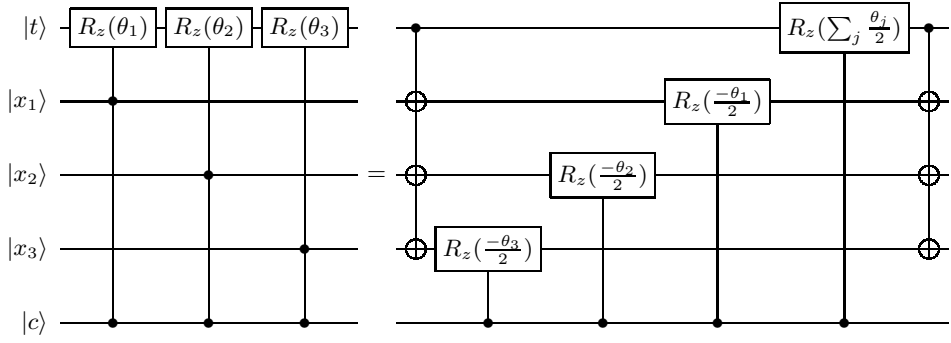
\begin{figure*}
    \centering
    \begin{minipage}{0.9\textwidth}
    \Qcircuit @C=0.4em @R=1em @!R{
    \lstick{\ket{t}} & \gate{R_z(\theta_1)} & \gate{R_z(\theta_2)} & \gate{R_z(\theta_3)} & \qw &&& \ctrl{1} & \qw & \qw & \qw & \gate{R_z(\sum_j \frac{\theta_j}{2})} & \ctrl{1} & \qw \\
    \lstick{\ket{x_1}} & \ctrl{-1} & \qw & \qw & \qw &&& \targ \qwx[1] & \qw & \qw & \gate{R_z(\frac{-\theta _1}{2})} & \qw & \targ \qwx[1] & \qw \\
    \lstick{\ket{x_2}} & \qw & \ctrl{-2} & \qw & \qw & \push{=} && \targ \qwx[1] & \qw & \gate{R_z(\frac{-\theta _2}{2})} & \qw & \qw & \targ \qwx[1] & \qw \\
    \lstick{\ket{x_3}} & \qw & \qw & \ctrl{-3} & \qw &&& \targ & \gate{R_z(\frac{-\theta _3}{2})} & \qw & \qw & \qw & \targ & \qw \\
    \lstick{\ket{c}} & \ctrl{-3} & \ctrl{-2} & \ctrl{-1} & \qw &&& \qw & \ctrl{-1} & \ctrl{-2} & \ctrl{-3} & \ctrl{-4} & \qw & \qw \\
    }
    \end{minipage}
    \caption{An illustration of a parallel circuit construction for shared controlled $R_z(\theta_j)$ gates.}
    \label{fig:CCnRz}
\end{figure*}
    Note that $Ry(\theta) = S H R_z(\theta) H S^\dagger$. Therefore, we add four single-qubit gates to the target qubit $\ket{t}$, reducing the problem to implementing the following transformation:
    \begin{align*}
        &\ket{t} \ket{x_1, \dots, x_n} \ket{c} \to R_z\left(c \sum_{j=1}^n x_j \theta_j\right) \ket{t} \ket{x_1, \dots, x_n} \ket{c}, \\
        &x_1, \dots, x_n \in \{0, 1\}, \quad c \in \{0, 1\},
    \end{align*}
    where $R_z(\theta)$ is defined as:
    \begin{equation*}
        R_z(\theta) = \begin{pmatrix}
            e^{-i\theta/2} & 0 \\
            0 & e^{i\theta/2}
        \end{pmatrix}.
    \end{equation*}

    Let the target qubit be in the state $\ket{t} = \alpha \ket{0} + \beta \ket{1}$, where $\alpha, \beta \in \mathbb{C}$ and $|\alpha|^2 + |\beta|^2 = 1$. Initially, the state is:
    \begin{equation*}
        (\alpha \ket{0} + \beta \ket{1}) \ket{x_1, \dots, x_n} \ket{c}.
    \end{equation*}
    We refer to the $n$ qubits in state $\ket{x_1, \dots, x_n}$ as input qubits, with the $j$-th input qubit in state $\ket{x_j}$. The control qubit is in state $\ket{c}$.

    \textbf{Step 1:} Use a quantum fan-out gate to propagate the target qubit state to the input qubits. By Theorem~\ref{the:fanout}, this requires a constant-depth circuit with $O(n)$ ancilla. The resulting state is:
    \begin{equation*}
        \alpha \ket{0} \ket{x_1, \dots, x_n} \ket{c} + \beta \ket{1} \ket{x_1 \oplus 1, \dots, x_n \oplus 1} \ket{c}.
    \end{equation*}

    \textbf{Step 2:} For each $j = 1, 2, \dots, n$, apply a controlled $R_z(-\theta_j/2)$ gate. The gate is controlled by the control qubit $\ket{c}$ and acts on the $j$-th input qubit. First, use Lemma~\ref{lem:GHZ} to copy the control qubit sufficiently. Then, apply all controlled $R_z$ gates in a single layer. Finally, recover the ancilla using Lemma~\ref{lem:recover}. This step requires a constant-depth circuit with $O(n)$ ancilla and yields the state:
    \begin{align*}
        &e^{i\frac{c}{4}\sum_{j=1}^n \theta_j (-1)^{x_j}} \alpha \ket{0} \ket{x_1, \dots, x_n} \ket{c} \\
        &+ e^{i\frac{c}{4}\sum_{j=1}^n \theta_j (-1)^{x_j \oplus 1}} \beta \ket{1} \ket{x_1 \oplus 1, \dots, x_n \oplus 1} \ket{c}.
    \end{align*}
    Since $x_j \in \{0, 1\}$, the state can be rewritten as:
    \begin{align*}
        &e^{i\frac{c}{4}\sum_{j=1}^n \theta_j (1 - 2x_j)} \alpha \ket{0} \ket{x_1, \dots, x_n} \ket{c} \\
        &+ e^{i\frac{c}{4}\sum_{j=1}^n \theta_j (2x_j - 1)} \beta \ket{1} \ket{x_1 \oplus 1, \dots, x_n \oplus 1} \ket{c}.
    \end{align*}

    \textbf{Step 3:} Apply the $R_z(\sum_j \theta_j/2)$ gate to the target qubit, resulting in:
    \begin{align*}
        &e^{i\frac{-c}{2}\sum_{j=1}^n \theta_j x_j} \alpha \ket{0} \ket{x_1, \dots, x_n} \ket{c} \\
        &+ e^{i\frac{c}{2}\sum_{j=1}^n \theta_j x_j} \beta \ket{1} \ket{x_1 \oplus 1, \dots, x_n \oplus 1} \ket{c}.
    \end{align*}

    \textbf{Step 4:} Repeat Step 1 to recover the original input qubit states, yielding the final state:
    \begin{equation*}
        \left( e^{i\frac{c}{2}\sum_{j=1}^n \theta_j x_j} \alpha \ket{0} + e^{i\frac{c}{2}\sum_{j=1}^n \theta_j x_j} \beta \ket{1} \right) \ket{x_1, \dots, x_n} \ket{c}.
    \end{equation*}
    This state is equivalent to the result of applying $R_z(c \sum_{j=1}^n x_j \theta_j)$. The entire circuit has constant depth, uses $O(n)$ ancilla, and requires four layers of intermediate measurements. An illustration for $n = 3$ is shown in Fig.~\ref{fig:CCnRz}. This completes the proof.
\end{proof}

\begin{theorem}
\label{the:cstatep}
    Given arbitrary \( n \)-qubit states \( \ket{\psi_j} \) for \( j = 0, 1, \dots, 2^n - 1 \), we can construct a quantum circuit \( C \) such that for an arbitrary \( n \)-qubit state \( \sum_{j = 0}^{2^n - 1} \alpha_j \ket{j} \), where \( \alpha_1, \dots, \alpha_{2^n - 1} \in \mathbb{C} \), the following holds:
\begin{equation*}
    C \left( \sum_{j=0}^{2^n-1} \alpha_j \ket{j} \otimes \ket{0}^{\otimes n} \right) = \sum_{j=0}^{2^n-1} \alpha_j \ket{j} \otimes \ket{\psi_j}.
\end{equation*}
The circuit \( C \) has a constant depth, a size of \( O(n4^n) \), utilizes \( O(n4^n) \) ancilla, and requires \( 50 \) layers of intermediate measurements.
\end{theorem}

\begin{proof}
    The proof proceeds in four steps:

    \textbf{Step 1:} Using Lemma~\ref{lem:toonehot}, we transform the initial state into:
    \begin{equation*}
        \sum_{j=0}^{2^n-1} \alpha_j \ket{e_j},
    \end{equation*}
    where $\ket{e_j}$ is the one-hot encoding of $j$. This step requires a constant-depth circuit with $O(n2^n \log n)$ ancilla and $9$ layers of intermediate measurements.

    \textbf{Step 2:} We modify Lemma~\ref{lem:onehot} to prepare the states $\ket{\psi_j}$ in parallel. Specifically:
    \begin{itemize}
        \item Steps 2 and 5 of Lemma~\ref{lem:onehot} remain unchanged, as they cancel each other out.
        \item Steps 1 and 3 are modified by replacing each $Ry$ gate with $2^n$ controlled $Ry$ gates, each controlled by a different qubit. By Lemma~\ref{lem:nRy}, this can be implemented in constant depth with $O(4^n)$ ancilla.
        \item Step 4 is modified by replacing each controlled $Ry$ gate with $2^n$ gates, as described in Lemma~\ref{lem:CCnRy}, requiring $O(4^n)$ ancilla.
        \item Step 6 remains unchanged, as only one condition for controlled state preparation is satisfied.
    \end{itemize}
    Let $one_{nophase}(\ket{\psi_j})$ denote the one-hot encoding of $\ket{\psi_j}$ without phases. For example, $one_{nophase}(\sum_{j=0}^{2^n-1} \alpha_j \ket{j}) = \sum_{j=0}^{2^n-1} |\alpha_j| \ket{e_j}$. After this step, the state becomes:
    \begin{equation*}
        \sum_{j=0}^{2^n-1} \alpha_j \ket{e_j} one_{nophase}(\ket{\psi_j}).
    \end{equation*}
    This step requires $30$ layers of intermediate measurements.

    \textbf{Step 3:} We add the phases to the state. In Theorem~\ref{the:statepre}, the $Z(\theta_j)$ gates are replaced with $2^n$ controlled $Z(\theta)$ gates, each controlled by one of the $2^n$ qubits. By Lemma~\ref{lem:nZ}, this step can be implemented in constant depth with $O(4^n)$ ancilla. Let $one(\ket{\psi_j})$ denote the one-hot encoding of $\ket{\psi_j}$ with phases. For example, $one(\sum_{j=0}^{2^n-1} \alpha_j \ket{j}) = \sum_{j=0}^{2^n-1} \alpha_j \ket{e_j}$. After this step, the state becomes:
    \begin{equation*}
        \sum_{j=0}^{2^n-1} \alpha_j \ket{e_j} one(\ket{\psi_j}).
    \end{equation*}
    This step requires $2$ layers of intermediate measurements.

    \textbf{Step 4:} We apply Lemma~\ref{lem:tobinary} to both the first $2^n$ qubits and the last $2^n$ qubits simultaneously. This step requires $9$ layers of intermediate measurements and yields the desired state:
    \begin{equation*}
        \sum_{j=0}^{2^n-1} \alpha_j \ket{j} \ket{\psi_j}.
    \end{equation*}

    The entire circuit has constant depth, uses $O(n4^n)$ ancilla, and requires $50$ layers of intermediate measurements. Since the circuit depth is constant, the circuit size is bounded by \( O(n + m) \) when the circuit utilizes \( m \) ancilla.
\end{proof}

\begin{corollary}
\label{cor:unitary}
    Given an arbitrary \( 2^n \times 2^n \) unitary \( U \), we can construct a constant-depth quantum circuit \( C \) with a size of \( O(n4^n) \). This circuit utilizes \( O(n4^n) \) ancilla and requires \( 50 \) layers of intermediate measurements. For an arbitrary \( n \)-qubit state \( \sum_{j = 0}^{2^n - 1} \alpha_j \ket{j} \), where \( \alpha_1, \dots, \alpha_{2^n - 1} \in \mathbb{C} \), the following holds:
    \begin{equation*}
        C \left( \sum_{j=0}^{2^n-1} \alpha_j \ket{j} \otimes \ket{0}^{\otimes n} \right) = \sum_{j=0}^{2^n-1} \alpha_j \ket{j} \otimes U\ket{j}.
    \end{equation*}
\end{corollary}
\begin{proof}
    Since $U\ket{j}$ is an $n$-qubit state for each $j = 0, 1, \dots, 2^n - 1$, this corollary follows directly from Theorem~\ref{the:cstatep}.
\end{proof}

\begin{corollary}
\label{cor:statepre}
    An arbitrary \( n \)-qubit quantum state can be prepared from the initial state \( \ket{0}^{\otimes n} \) using a constant-depth quantum circuit with a size of \( O(n2^n) \). This circuit utilizes \( O(n2^n) \) ancilla and requires \( 81 \) layers of intermediate measurements.
\end{corollary}
\begin{proof}
    For an arbitrary \( n \)-qubit state \( \sum_{j=0}^{2^n-1} \alpha_j \ket{j} \), where \( \alpha_0, \dots, \alpha_{2^n-1} \in \mathbb{R} \) and \( \sum_{j=0}^{2^n-1} |\alpha_j|^2 = 1 \), Theorem~\ref{the:statepre} guarantees that this state can be prepared using a constant-depth circuit with \( O(n4^n) \) ancilla.

For an arbitrary \( n \)-qubit state \( \sum_{k=0}^{2^n-1} \beta_{j,k} \ket{k} \), where \( j = 0, 1, \dots, 2^n - 1 \) and \( \beta_{j,k} \in \mathbb{C} \), Theorem~\ref{the:cstatep} ensures the existence of a constant-depth circuit \( C \) with \( O(n4^n) \) ancilla that satisfies:
\begin{equation*}
    C \left( \sum_{j=0}^{2^n-1} \alpha_j \ket{j} \otimes \ket{0}^{\otimes n} \right) = \sum_{j=0}^{2^n-1} \alpha_j \ket{j} \otimes \sum_{k=0}^{2^n-1} \beta_{j,k} \ket{k}.
\end{equation*}

Notably, any \( 2n \)-qubit state can be expressed in the form \( \sum_{j=0}^{2^n-1} \alpha_j \ket{j} \otimes \sum_{k=0}^{2^n-1} \beta_{j,k} \ket{k} \). Therefore, an arbitrary \( 2n \)-qubit state can be prepared using a constant-depth circuit with \( O(n4^n) \) ancilla. Similarly, for an arbitrary \( (2n - 1) \)-qubit state \( \ket{\phi} \), the state \( \ket{\phi} \otimes \ket{0} \) can be prepared using a constant-depth circuit with \( O(n4^n) \) ancilla. Consequently, an arbitrary \( n \)-qubit state can be prepared using a constant-depth circuit with \( O(n2^n) \) ancilla. Since the circuit depth is constant, the circuit size is bounded by the number of qubits \( O(n2^n) \). This circuit requires \( 81 \) layers of intermediate measurements, as established by Theorem~\ref{the:statepre} and Theorem~\ref{the:cstatep}.
\end{proof}

According to \cite{takahashi2016collapse}, the circuit size of \( O(n \log n) \) in Theorem~\ref{the:mtoffoli} can be reduced to \( O(n \log^{(c)} n) \) for any positive constant \( c \), while maintaining constant circuit depth. Here, \( \log^{(c)} n \) denotes the \( c \)-times iterated logarithm. In practice, by setting \( c = 5 \), the \( n \)-Toffoli gate can be implemented using a constant-depth circuit with a size of \( O(n) \), since \( \log^{(5)} n < 10 \) for all practical values of \( n \). Applying this technique, we eliminate the factor of \( n \) in the circuit sizes \( O(n2^n) \) and \( O(n4^n) \), reducing them to \( O(2^n) \) and \( O(4^n) \), respectively, as shown in Table~\ref{tab:result}. This reduction is achieved by optimizing the circuit size of Toffoli from \( O(n \log n) \) to \( O(n) \) in Steps 2 and 5 of Lemma~\ref{lem:onehot}, which serve as fundamental building blocks in our circuit construction.

\section{Constant-Depth Circuit for Reversible Functions}
\label{app:refunction}

\begin{definition}
    A function \( f: \{0,1\}^n \to \{0,1\}^n \) is \textit{reversible} if and only if for any \(\boldsymbol{x}, \boldsymbol{y} \in \{0,1\}^n\) where \(\boldsymbol{x} \neq \boldsymbol{y}\), it holds that \( f(\boldsymbol{x}) \neq f(\boldsymbol{y}) \).
\end{definition}

\begin{theorem}
    For any reversible function \( f: \{0,1\}^n \to \{0,1\}^n \), there exists a constant-depth quantum circuit \( C \) with \( O(n2^n \log n) \) ancilla that implements \( f \) as follows. This circuit requires \( 18 \) layers of intermediate measurements:
    \begin{equation*}
        C \ket{\boldsymbol{x}} = \ket{f(\boldsymbol{x})}, \quad \boldsymbol{x} \in \{0,1\}^n.
    \end{equation*}
\end{theorem}

\begin{proof}
    The proof proceeds in three steps:

    \textbf{Step 1:} Convert the binary representation of the input state into its one-hot representation. Let the initial state be:
    \begin{equation*}
        \sum_{\boldsymbol{x} \in \{0,1\}^n} \alpha_{\boldsymbol{x}} \ket{\boldsymbol{x}}, \quad \alpha_{\boldsymbol{x}} \in \mathbb{C}.
    \end{equation*}
    Using Lemma~\ref{lem:toonehot}, we transform this state into:
    \begin{equation*}
        \sum_{\boldsymbol{x} \in \{0,1\}^n} \alpha_{\boldsymbol{x}} \ket{e_{\boldsymbol{x}}},
    \end{equation*}
    where \(\ket{e_{\boldsymbol{x}}}\) is the one-hot encoding of \(\boldsymbol{x}\). This step requires a constant-depth circuit with \( O(n2^n \log n) \) ancilla and \( 9 \) layers of intermediate measurements.

    \textbf{Step 2:} Transform the one-hot encoded state \(\ket{e_{\boldsymbol{x}}}\) into \(\ket{e_{f(\boldsymbol{x})}}\). Since \(\ket{e_{\boldsymbol{x}}}\) has exactly one qubit in state \(\ket{1}\), we can use a circuit composed solely of swap gates to map \(\ket{e_{\boldsymbol{x}}}\) to \(\ket{e_{f(\boldsymbol{x})}}\). By Proposition 1 in \cite{moore2001parallel}, such a swap-based circuit can be implemented using \( 6 \) layers of CNOT gates without additional ancilla.

    \textbf{Step 3:} Convert the one-hot encoded state \(\ket{e_{f(\boldsymbol{x})}}\) back into the binary representation. Using Lemma~\ref{lem:tobinary}, we obtain the final state:
    \begin{equation*}
        \sum_{\boldsymbol{x} \in \{0,1\}^n} \alpha_{\boldsymbol{x}} \ket{f(\boldsymbol{x})}.
    \end{equation*}
    This step requires a constant-depth circuit with \( O(n2^n \log n) \) ancilla and \( 9 \) layers of intermediate measurements.

    The entire circuit has constant depth, uses \( O(n2^n \log n) \) ancilla, and requires \( 18 \) layers of intermediate measurements. This completes the proof.
\end{proof}

% \bibliography{apssamp}% Produces the bibliography via BibTeX.
\bibliographystyle{IEEEtran}
\bibliography{ref}

\end{document}